\documentclass[preprint,12pt]{elsarticle}

\usepackage{graphicx}
\usepackage{svg}
\usepackage{amssymb}
\usepackage{amsmath}
\usepackage{adjustbox}
\usepackage{xcolor}
\usepackage{colortbl}
\usepackage{tabularx}
\usepackage{booktabs}
\usepackage{multirow}
\usepackage{url}
\usepackage{bbding}
\usepackage{tikz}
\usetikzlibrary{
    fit,shapes.misc
}

\usepackage{geometry}
\usepackage{color}
\usepackage{fancyvrb}
\usepackage{hyperref}
\hypersetup{
    colorlinks=true,
    linkcolor=blue,
    urlcolor=blue,
}
\usepackage{textcomp} 

\usepackage{draftwatermark}
\SetWatermarkText{PREPRINT}
\SetWatermarkScale{3}

\usepackage{makecell}

\usepackage{easyReview}
\def\remove#1{}

\newcolumntype{?}{!{\vrule width 1pt}}

\journal{Applied Soft Computing}

\begin{document}

\pagenumbering{gobble}
\setcounter{page}{0}

\section*{Elsevier Copyright Notice}

\noindent
Copyright \textcopyright $\;$ [2025] Elsevier. This is a post-peer-review version of an article published in \textit{Applied Soft Computing}. This manuscript is made available under the Elsevier user license. 

\vspace{3em}
\noindent\textbf{Published in:} Applied Soft Computing, 2025.

\vspace{3em}
\noindent\textbf{Cite as:}
\vspace{0.5em}

\noindent
Pedro R. Pires, Tiago A. Almeida, Interact2Vec -- An efficient neural network-based model for simultaneously learning users and items embeddings in recommender systems, Applied Soft Computing, 181 (2025) 113408:1--113408:17, http://dx.doi.org/ 10.1016/j.asoc.2025.113408.

\vspace{3em}
\noindent\textbf{BibTeX:}
\vspace{0.5em}

\begin{Verbatim}[frame=single]
@article{PIRES2025113408,
    title    = {Interact2Vec -- An efficient neural network-based
                model for simultaneously learning users and items
                embeddings in recommender systems},
    journal  = {Applied Soft Computing},
    volume   = {181},
    pages    = {113408:1--113408:17},
    year     = {2025},
    issn     = {1568-4946},
    doi      = {https://doi.org/10.1016/j.asoc.2025.113408},
    url      = {https://www.sciencedirect.com/science/article/pii/
                S1568494625007197},
    author   = {Pedro R. Pires and Tiago A. Almeida},
    keywords = {Recommender systems, Collaborative filtering, 
                Distributed vector representation, Embeddings},
}
\end{Verbatim}

\newpage
\pagenumbering{arabic}

\begin{frontmatter}

\title{Interact2Vec -- An efficient neural network-based model for simultaneously learning users and items embeddings in recommender systems}

\author[end1]{Pedro R. Pires\corref{mycorrespondingauthor}}
\address[end1]{Department of Computer Science, Federal University of S\~{a}o Carlos, \\Sorocaba, S\~{a}o Paulo, 18052-780, Brazil}

\cortext[mycorrespondingauthor]{Corresponding author}
\ead{pedro.pires@dcomp.sor.ufscar.br}

\author[end1]{Tiago A. Almeida}


\begin{abstract}
Over the past decade, recommender systems have experienced a surge in popularity. Despite notable progress, they grapple with challenging issues, such as high data dimensionality and sparseness. Representing users and items as low-dimensional embeddings learned via neural networks has become a leading solution. However, while recent studies show promising results, many approaches rely on complex architectures or require content data, which may not always be available. This paper presents Interact2Vec, a novel neural network-based model that simultaneously learns distributed embeddings for users and items while demanding only implicit feedback. The model employs state-of-the-art strategies that natural language processing models commonly use to optimize the training phase and enhance the final embeddings. Two types of experiments were conducted regarding the extrinsic and intrinsic quality of the model. In the former, we benchmarked the recommendations generated by Interact2Vec's embeddings in a top-$N$ ranking problem, comparing them with six other recommender algorithms. The model achieved the second or third-best results in 30\% of the datasets, being competitive with other recommenders, and has proven to be very efficient with an average training time reduction of 274\% compared to other embedding-based models. Later, we analyzed the intrinsic quality of the embeddings through similarity tables. Our findings suggest that Interact2Vec can achieve promising results, especially on the extrinsic task, and is an excellent embedding-generator model for scenarios of scarce computing resources, enabling the learning of item and user embeddings simultaneously and efficiently.
\end{abstract}



\begin{keyword}
recommender systems \sep collaborative filtering \sep distributed vector representation \sep embeddings
\end{keyword}

\end{frontmatter}

\baselineskip=0.97\baselineskip

\section{Introduction}\label{sec:introduction}
As technology advances and content becomes increasingly accessible, a growing volume of data is generated and shared daily. While this has led to numerous advancements in the modern world, the sheer magnitude of information means that only a fraction is relevant to individual users. To alleviate this problem, Recommender Systems (RS) emerged in the 1990s and have progressively become integral parts of our digital lives~\cite{Bobadilla2013}. Collaborative Filtering (CF) stands out as the most popular among different types of RS. Its high popularity can be credited to its simplicity since it requires only a matrix of user-item interactions to be trained, data easily obtained in recommender systems scenarios~\cite{Bobadilla2013}.

The first CF recommender systems used the concept of neighborhood and similarity to generate recommendations. In user-based CF, users are represented as binary vectors of consumed items and receive recommendations according to the items consumed by other users in their neighborhood. This neighborhood is constructed through similarity metrics between the users' vectors~\cite{Herlocker2002}. Similarly, in item-based CF, the recommendation can be generated using similarity between items, represented as vectors of the users who have consumed them. Although this type of representation is capable of generating good recommendations, the accelerated development of RS resulted in problems of sparsity and scalability, given that modern RS has to deal with an ever-growing number of items and users but with a much smaller relative fraction of interactions~\cite{Khusro2016}. Numerous studies have suggested the representation of users and items in a reduced dimensional space as a solution to overcome these challenges. In this context, matrix factorization algorithms have become popular, using matrix decomposition techniques to learn low-dimensional arrays of latent factors for users and items~\cite{Koren2009a}.

Recent studies in the RS literature proposed using neural models inspired by state-of-the-art Natural Language Processing (NLP) techniques~\cite{Mikolov2013b}, capable of learning user and item embeddings, i.e., dense vectors with low and fixed dimensions that carry intrinsic meaning. Offering a solution to scalability and sparsity challenges, neural embeddings have become an appealing option in recommendation problems. The first NLP-based neural embedding models in the context of recommendation were Prod2Vec, User2Vec~\cite{Grbovic2015}, and Item2Vec~\cite{Barkan2016}. Those models are adaptations of neural networks commonly used to learn word embeddings and achieve results comparable to established algorithms. 

However, despite numerous advancements and proposals regarding this strategy, recent approaches often consist of training complex neural networks or consuming additional content. Many state-of-the-art embedding-based neural networks, such as SASRec~\cite{Kang2018}, BERT4Rec~\cite{Sun2019}, and CL4SRec~\cite{Xie2022}, demand learning through sessions and sequential interactions of the user with the system. Other approaches, such as Attr2Vec~\cite{Fu2017} and Hasanzadeh \textit{et al.}'s~\cite{Hasanzadeh2022} review-based recommender, demand item metadata, which is often unreliable and not readily available in every recommender system. Additionally, many embedding-based algorithms still focus on the problem of rating prediction with explicit feedback, such as RecDNNing~\cite{Zarzour2019}, which can compromise its application since it is a piece of information often unavailable. Finally, since many state-of-the-art algorithms tend to use complex neural networks, processing power can be a problem in settings with limited resources.

This paper builds upon recent studies on applying distributed vector representation models learned by neural networks in collaborative filtering recommender systems. To provide a straightforward application approach that demands low computational resources, we introduce Interact2Vec, a method to jointly and efficiently generate vector representations of items and users in the same low-dimensional vector space, which can then be used for top-$N$ recommendation. We analyzed the model both extrinsically and intrinsically, with a top-$N$ ranking task and similarity tables, respectively. Results show how the primary strength of Interact2Vec lies in its high efficiency while being proven statistically superior to User2Vec and having surpassed Item2Vec training time considerably in many different datasets. This shows how the model can be applied to scenarios with limited computational resources while maintaining a competitive performance compared to other recommenders.

\section{Related work}\label{sec:related_work}
Collaborative filtering is currently the most widely used technique in recommender systems, given that it requires only an easily obtainable matrix of user-item interactions to generate recommendations~\cite{Bobadilla2013}. In pioneer approaches, neighborhood-based algorithms figured as one of the most traditional techniques~\cite{Herlocker2002}. However, due to scalability and sparsity issues~\cite{Khusro2016}, they have been replaced by matrix factorization-based methods, that quickly became state-of-the-art~\cite{Koren2009a}.

While matrix factorization methods are known for their high quality, their use is limited in sparse datasets, and they are computationally expensive~\cite{Zhang2016a, Bobadilla2013}. Recently, new approaches inspired by state-of-the-art Natural Language Processing (NLP) techniques have emerged to tackle these issues by learning neural embeddings for items or users. Those neural embeddings are dense vector representations of low and fixed dimensionality, which carry intrinsic meaning and are computed by artificial neural networks.

With the success of Word2Vec, neural models for creating context-based word embeddings garnered significant attention in the field of NLP. They proved to be effective without requiring a lot of computing work~\cite{Mikolov2013b}. Currently, neural embedding is one of the primary methods for solving numerous issues in domains other than NLP~\cite{Camacho2018}. 

In the recommender system area, one of the earliest uses of neural models to generate item embeddings is the Prod2Vec~\cite{Grbovic2015}. The model was heavily inspired by Skip-gram~\cite{Mikolov2013b}, achieving higher accuracy than other heuristic recommenders used as a baseline in a recommendation scenario via e-mail. In the same study, the authors also presented User2Vec, the first approach to compute user and item neural embeddings concomitantly by feeding the model with a composition of the vector representation of the user and certain consumed products.

In the following year, Item2Vec was proposed~\cite{Barkan2016}, an adaptation of Skip-gram's architecture with a reformulated concept of context: any two items purchased by the same user are related, thus using a variable size window to capture correlated items. Item2Vec was superior to matrix factorization models in several intrinsic evaluation tasks.

Prod2Vec and Item2Vec are pure collaborative filtering methods based on implicit feedback, i.e., the models learn only with the consumption of user-item implicit interactions. The subsequent research sought to consider a hybrid filtering scenario with the consumption of descriptive content to enrich the embeddings, adapting the previous embedding-based recommenders~\cite{Fu2017,Vasile2016} or feeding item metadata to different neural architectures~\cite{Greenstein2017,Zhang2016a}.

Other studies employed more complex neural networks, with a dominant strategy converting the recommendation problem to a sequence forecasting problem and applying convolutional~\cite{Tang2018}, recurrent~\cite{Hidasi2016,Tan2016}, or attention-based neural networks~\cite{Kang2018,Sun2019,Xie2022}. Deep learning models were also used to learn the embeddings~\cite{Sidana2021,Zarzour2019}. Although new proposals have suggested using complex neural architectures, much of the recent research focuses on training simpler models inspired by Word2Vec~\cite{Grbovic2018,Valcarce2019} or applying NLP models directly over textual information of items~\cite{Collins2019,Hasanzadeh2022}.

Finally, one recent strategy for learning user and item embeddings consists of using graph-based techniques, a straightforward approach due to the relational aspect of the recommendation problem~\cite{Deng2022}. This technique is used mainly in the context of social recommenders~\cite{Liu2019} and bundle recommendations~\cite{Wei2023} and has gained attention with the development of heterogeneous information networks (HIN)~\cite{Liu2022}. HIN-based recommender systems represent user and items as nodes on a graph and their relations as edges. Instead of homogeneous approaches, they use auxiliary information to enrich the graph's entities. This can be done using metadata for the items~\cite{Yu2013,Pham2023} or the relations itself~\cite{Forouzandeh2024}. However, many open problems can still compromise the final representation, such as \textit{(i)} scalability issues, as graph embeddings for large-scale graphs can be computationally expensive to calculate, \textit{(ii) }demand for additional information since HIN-based recommenders demand auxiliary data to build the model, and \textit{(iii)} complexity in maintaining embedding quality since the representation of dynamic scenarios is constructed with static graphs~\cite{Gao2023}. Although recent studies have proposed graph-based embedding models that are efficient and scalable, such as LightGCN~\cite{He2020}, this study focused on item and user-embedding-based models that are trained through traditional neural network approaches.

The literature shows that most recent studies have focused on three different strategies for improving item and user embeddings: \textit{(i)} using content information to enrich the representation, which, although it can achieve competitive results, has the disadvantage of not being easily applicable since they depend on additional data that may not be available;  \textit{(ii)} employing neural models exclusively to generate item embeddings, ignoring user embeddings or learning them through simple heuristics, as a traditional averaging of item embeddings which can result in loss of information and impact the quality of the model~\cite{Bendada2023}; and \textit{(iii)} using very complex neural models or graph representation, which are difficult to apply in scenarios where computational resources are scarce and can even generate a false impression of better performance when evaluated in the wrong manner~\cite{Dacrema2019}.

In addition, few studies on neural embeddings have analyzed the intrinsic meaning carried by the vector representation~\cite{Gladkova2016}. Normally, the embeddings are evaluated in downstream applications, i.e., the recommendation itself, which is not a reliable way to intrinsically access the representation~\cite{Schnabel2015}. Since embeddings can be used in other tasks, such as auto-tagging, knowledge discovery, user-item clustering, and the discovery of categorical features~\cite{Chang2023}, evaluating how well they represent content-based information is important.

Many strategies can be adopted when evaluating the intrinsic quality of the vector representations, such as genre plotting~\cite{Barkan2016,Fu2017,Siswanto2018} and analogy analysis~\cite{Greenstein2017}. However, from the different existing procedures, similarity tables are the most commonly used approach due to their simplicity and ease of interpretation~\cite{Barkan2016,Fu2017,Grbovic2015}.

Inspired by other collaborative filtering shallow neural networks that require small computational effort, we propose a new neural model for learning embeddings in recommender systems, which we named Interact2Vec. Although computationally more efficient, our model has a similar network architecture to Item2Vec but a similar objective to User2Vec, simultaneously learning embeddings of items and users. We then compare Interact2Vec with different state-of-the-art models that share simplicity-related similarities. Results in extrinsic and intrinsic tasks show how our proposed model can achieve competing results with other recommenders while being computationally efficient and easy to apply.

\section{Interact2Vec}\label{interact2vec}

The Interact2Vec is a novel neural model that simultaneously generates user and item embeddings. We have designed it to meet the objectives and hypotheses described in Table~\ref{tab:int2v_objectives}. With that in mind, the model consists of a shallow neural network, computationally efficient, that requires only the user-item interaction matrix to be trained.

\begin{table}[htpb]
    \centering
    \begin{tabular}{@{}cc@{}}\toprule
         \thead{\textbf{Objective}} & \thead{\textbf{Hypothesis}} \\ \midrule
         \textbf{\makecell{Low dimensionality \\ and sparsity}} & \makecell{The representation learned by the model \\ must have low dimensionality and \\ be dense, thus avoiding the \\ high sparsity problem} \\ \hline
        \textbf{\makecell{Computational \\ efficiency}} & \makecell{The model must present a \\ computational cost equal to or lower than \\ other recommendation methods \\ based on vector representations} \\ \hline
        \textbf{Ease of application} & \makecell{The model should only require an array of \\ implicit feedback interactions to be trained, \\ without the need for content data \\ or additional information} \\ \hline
    \end{tabular}
    \caption{Objectives and hypotheses of Interact2Vec}
    \label{tab:int2v_objectives}
\end{table}

The following subsections explain how Interact2Vec works, its neural architecture and objective function, techniques used during the learning phase, and different strategies for yielding recommendations using the embeddings. Ultimately, we analyze its computational complexity compared with the baseline embedding-based models.

\subsection{Neural architecture and objective function}

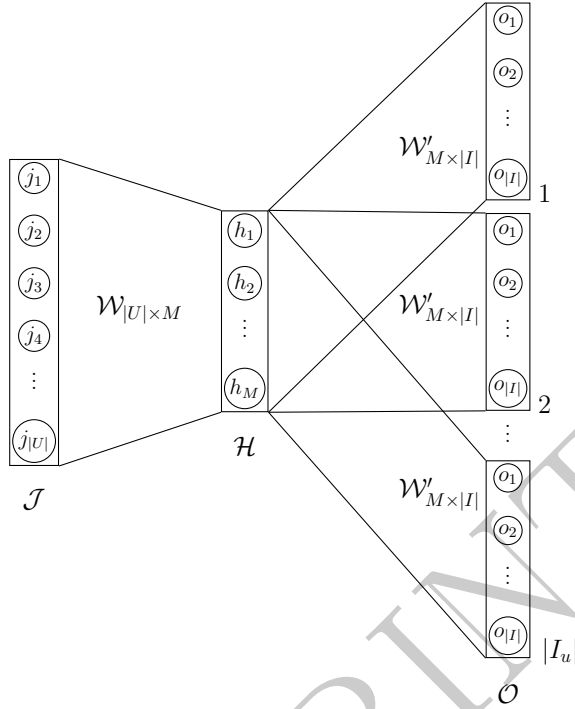
\begin{figure}[htb!]
\centering  
{
\tikzset{
group/.style={rectangle, draw, inner sep=1pt, minimum size=3mm,transform shape=false},
input/.style={circle, draw, inner sep=1pt, minimum size=3mm},
missing/.style={draw=none,text height=1,
    execute at begin node=\color{black}$\vdots$},
hidden/.style={circle, draw, inner sep=1pt, minimum size=3mm},
output/.style={circle, draw, inner sep=1pt, minimum size=3mm},
el/.style={pos=0.2, fill=white, font=\sffamily\fontsize{4}{4},inner sep=0pt},
tx/.style={font=\bf\sffamily\fontsize{14}{14}}
}
\begin{tikzpicture}[node distance=1cm,scale=0.7,transform shape]]
  \node[input] (x1) at (0,-6)  {$j_1$};
  \node[input] (x2) at (0,-7)  {$j_2$};
  \node[input] (x3) at (0,-8)  {$j_3$};
  \node[input] (x4) at (0,-9)  {$j_4$};
  \node[missing] (mx) at (0,-10)  {};
  \node[input] (xn) at (0,-11)  {$j_{|U|}$};
  \node[group, fit={(x1) (x2) (x3) (x4) (xn)}] (ingroup) {};
  
  \node[hidden] (h1) at (4,-7)  {$h_1$};
  \node[hidden] (h2) at (4,-8)  {$h_2$};
  \node[missing] (mh) at (4,-9)  {};
  \node[hidden] (hm) at (4,-10)  {$h_M$};
  \node[group, fit={(h1) (h2) (hm)}] (hiddengroup) {};

  \node[output] (y11) at (9,-3)  {$o_1$};
  \node[output] (y12) at (9,-4)  {$o_2$};
  \node[missing] (my1) at (9,-5)  {};
  \node[output] (y1n) at (9,-6)  {$o_{|I|}$};
  \node[group, fit={(y11) (y12) (y1n)}] (outgroup1) {};

  \node[output] (y21) at (9,-7)  {$o_1$};
  \node[output] (y22) at (9,-8)  {$o_2$};
  \node[missing] (my2) at (9,-9)  {};
  \node[output] (y2n) at (9,-10)  {$o_{|I|}$};
  \node[group, fit={(y21) (y22) (y2n)}] (outgroup2) {};

  \node[output] (y31) at (9,-11.7)  {$o_1$};
  \node[output] (y32) at (9,-12.7)  {$o_2$};
  \node[missing] (my3) at (9,-13.7)  {};
  \node[output] (y3n) at (9,-14.7)  {$o_{|I|}$};
  \node[group, fit={(y31) (y32) (y3n)}] (outgroup3) {};
  
  \node[tx] at (2,-8.5) {$\mathcal{W}_{{|U|} \times M}$};
  \node[tx] at (7.7,-5.5) {$\mathcal{W}'_{M \times {|I|}}$};
  \node[tx] at (7.7,-8.5) {$\mathcal{W}'_{M \times {|I|}}$};
  \node[tx] at (7.7,-12) {$\mathcal{W}'_{M \times {|I|}}$};
  \node[tx] at (9.7,-6.3) {$1$};
  \node[tx] at (9.7,-10.3) {$2$};
  \node[missing] at (9,-11) {};
  \node[tx] at (10,-15) {$|I_u|$};
  
  \node[tx] at (0,-12.1) {$\mathcal{J}$};
  \node[tx] at (4,-11.1) {$\mathcal{H}$};
  \node[tx] at (9,-15.8) {$\mathcal{O}$};
  
  \draw(outgroup1.north west)--(hiddengroup.north east);
  \draw(outgroup1.south west)--(hiddengroup.south east);
  \draw(outgroup2.north west)--(hiddengroup.north east);
  \draw(outgroup2.south west)--(hiddengroup.south east);
  \draw(outgroup3.north west)--(hiddengroup.north east);
  \draw(outgroup3.south west)--(hiddengroup.south east);
  \draw(ingroup.north east)--(hiddengroup.north west);
  \draw(ingroup.south east)--(hiddengroup.south west);
\end{tikzpicture}
\caption{Interact2Vec's neural architecture}
\label{fig:in2v}
}
\end{figure}

Interact2Vec is a shallow artificial neural network with an architecture inspired by Item2Vec~\cite{Barkan2016} and an objective function inspired by User2Vec~\cite{Grbovic2015}. The network has three layers: an input layer $\mathcal{J}$, a hidden layer $\mathcal{H}$, and an output layer $\mathcal{O}$, as illustrated in Figure~\ref{fig:in2v}.

First, we will define a recommender system as a set $R$ of interactions between two main entities: a set $U$ of $|U|$ users and a set $I$ of $|I|$ items. $R$ comprises elements $r_{ui}$ representing the interaction between a user $u \mid u \in U$ and an item $i \mid i \in I $. Thus, if user $u$ has previously interacted with item $i$, we have that $r_{ui} \in R$; otherwise, $r_{ui} \notin R$. We will also denote the subset of items consumed by $u$ as $I_u = \{i \in I : r_{ui} \in R_u\}$, and the subset of users who consumed $i$ as $U_i = \{u \in U : r_{ui} \in R_i\}$. Consequently, $|I_u|$ represents the number of items consumed by user $u$, and  $|U_i|$ is the number of users who interacted with item $i$.

The input layer $\mathcal{J}$ consists of a vector of size $|U|$ representing a target user $u$ as a one-hot encoded vector. $\mathcal{H}$ has size $M$, corresponding to the number of dimensions desired for the final embeddings. It is connected to $\mathcal{J}$ by a weight matrix $\mathcal{W}$ of shape $|U| \times M$, thus $\mathcal{H} = \mathcal{J} \cdot \mathcal{W}$. Finally, $\mathcal{O}$ contains $|I_u|$ vectors of size $|I|$ (number of items), representing the items consumed by user $u$, also one-hot encoded. Each vector is connected to $\mathcal{H}$ by the same weight matrix $\mathcal{W}'$ of shape $M \times |I|$ and a softmax activation function. Thus $\mathcal{O} = softmax(\mathcal{H}^T \cdot \mathcal{W}')$.

Like User2Vec, Interact2Vec learns user and item embeddings by consuming user-item relationships. However, contrary to the other model, which trains the network by combining user and item embeddings as input, Interact2Vec's input layer is fed only with user embeddings. Given a user $u$ as input, its main objective is to predict all items $i$ previously consumed by $u$, i.e., $i \in I_u$, maximizing Equation~\ref{eq:n2v_objetivo}:

\begin{equation}\label{eq:n2v_objetivo}
    \frac{1}{|U|}\sum_{u \in U}\Big({\frac{1}{|I_u|}\sum_{i \in I_u}{\text{log } \sigma(u, i)}}\Big)
\end{equation}

\noindent in which $\sigma$ is the softmax activation function described in Equation~\ref{eq:softmax}:

\begin{equation}\label{eq:softmax}
    \sigma(u, i) = \frac{e^{(\mathcal{W}_u \mathcal{W}'^T_i)}}{\sum_{j \in I}{e^{(\mathcal{W}_u \mathcal{W}'^T_j)}}}
\end{equation}

\noindent in which $\mathcal{W}_u$ and $\mathcal{W}'^T_i$ contain user $u$ and item $i$ embeddings, respectively.

We compute the error as the sum of the differences between the expected one-hot vectors and the predicted outputs for each epoch, as in a traditional classification problem. We then update the network weight matrices by backpropagation. After training, the rows of $\mathcal{W}$ will contain the user embeddings, while the columns of $\mathcal{W}'$ will contain the item embeddings.

Interact2Vec's objective function is similar to Item2Vec's as they share a similar architecture. However, while in Item2Vec the model seeks to maximize its predictive capacity over pairs of items consumed by a user, in Interact2Vec, it maximizes its ability to predict a single item given a user, as it is commonly done in traditional matrix factorization methods~\cite{Hu2008,Koren2009a}.

\subsection{Simplified architecture}\label{sec:n2v_pratica}

The training phase of the Interact2Vec neural network can be simplified to a multiplication of the user and item embeddings for each user-item pair, updating the weights so that this operation returns 0 or 1, depending on whether the interaction occurs or not.

The multiplication of the input layer $\mathcal{J}$, containing a one-hot encoding of user $u$, by the weight matrix $\mathcal{W}$, will activate the row of $\mathcal{W}$ that refers to the current user $u$, i.e. its representation. For this reason, the intermediate layer $\mathcal{H}$ can be called a projection layer since it projects an embedding.

In the next step, the projected user embedding is multiplied by the embedding of all items stored in $\mathcal{W}'$. Considering that the desired outputs are the items consumed by the user, also represented in one-hot encoding, the network seeks to learn weights in such a way that:

\begin{equation}\label{eq:n2v_dot_prod}
    \mathcal{W}_u \times \mathcal{W}'^T_i = \begin{cases}
        1, & \text{if the user } u \text{ has interacted with item } i \\ 0, & \text{otherwise}
    \end{cases}
\end{equation}

\begin{figure}[htpb!]
\centering  
{
\tikzset{
input/.style={circle, draw, inner sep=12pt, minimum size=10mm}
}
\begin{tikzpicture}[node distance=1cm,scale=0.8,transform shape]]
  \node[input] (x1) at (-9,1)  {\LARGE{$u$}};
  \node[input] (x2) at (-9,-1)  {\LARGE{$i$}};
  
  \node[rectangle,draw,text width=3.5cm,align=center] (lkpu) at (-5.4,1) {Look-up table of user embeddings};
  \node[rectangle,draw,text width=3.5cm,align=center] (lkpi) at (-5.4,-1) {Look-up table of item embeddings};
  
  \node[rectangle,draw,text width=2cm,align=center] (wu) at (-1.6,1)  {\Large{$\mathcal{W}_u$}};
  \node[rectangle,draw,text width=2cm,align=center] (wi) at (-1.6,-1)  {\Large{$\mathcal{W}'^T_i$}};
  
  \node[input] (d1) at (1.55,0)  {\Large{$\times$}};
  \node[input] (s1) at (3.9,0)  {\Large{$\phi$}};
  \node[input] (o1) at (6.3,0)  {0/1};
  
  \draw(x1.east)--(lkpu.west);
  \draw(x2.east)--(lkpi.west);
  \draw(lkpu.east)--(wu.west);
  \draw(lkpi.east)--(wi.west);
  \draw(wu.east)--(d1.west);
  \draw(wi.east)--(d1.west);
  \draw(d1.east)--(s1.west);
  \draw(s1.east)--(o1.west);
\end{tikzpicture}
\caption{The practical architecture of Interact2Vec}
\label{fig:in2v_pratica}
}
\end{figure}
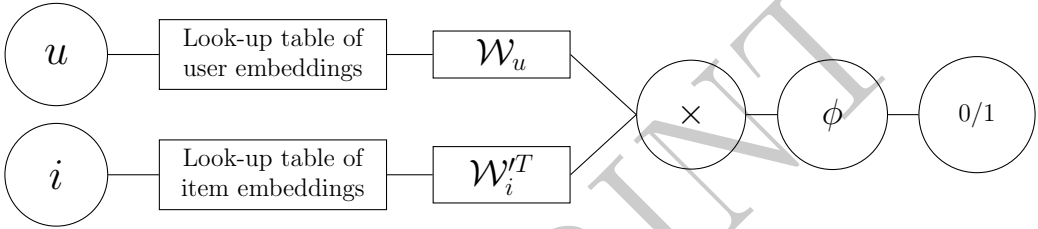

Thus, another way to interpret the model is: for each possible user-item pair, the network must \textit{(i)} retrieve the user and item embeddings by consulting look-up tables; \textit{(ii)} calculate its dot product, as specified by Equation~\ref{eq:n2v_dot_prod}; \textit{(iii)} normalize the obtained value between a 0 and 1 range, using a sigmoid activation function $\phi$  (Equation~\ref{eq:sigmoid}), and \textit{(iv)} update the content of the embeddings to approximate the obtained result closer to the desired one. The network architecture then becomes the one shown in Figure~\ref{fig:in2v_pratica}.

\begin{equation}\label{eq:sigmoid}
    \phi(u, i) = \frac{1}{1 + e^{-(\mathcal{W}_u \mathcal{W}'^T_j)}}
\end{equation}

Interpreting the architecture in this way can help simplify the model's implementation and enable the use of different traditional learning techniques in neural-embeddings methods, which will be explained in the following section.

\subsection{Strategies to optimize network training}

Like other neural embedding models based on NLP models, Interact2Vec employs several techniques during the learning stage to increase its generalization power and reduce its computational cost. 
This allows the method to better adapt to the application domain and require less processing power. 
In the following, we present each of these techniques.

\subsubsection{Subsampling of frequent items}\label{sec:freq_subsampling}

In NLP applications, performing a subsampling of frequent words before training can improve the generalization power of the model and its performance~\cite{Mikolov2013b}. In NLP tasks, this strategy defines a probability distribution for every word in the corpus, assigns higher probabilities for words that reoccur more frequently, and drops words from the sentence according to the calculated probability. In opposition to a simple stop-word removal\footnote{A data cleaning process that completely removes common words, such as \textit{``the''} and \textit{``you''}.}, the main goal of subsampling frequent words is to maintain their occurrence in the dataset while counterbalancing with the occurrence of less frequent and rarer words.

This strategy has been employed by multiple embedding-based recommender models~\cite{Barkan2016,Vasile2016,Caselles2018}. In the recommender systems scenario, the removal is conducted at the level of the interactions, reducing the number of occurrences of popular items, i.e., items that are recurrent within the dataset. A probability function for discarding the interaction of certain items is normally adopted. Before training the Interact2Vec, we employed the subsampling function presented in Equation~\ref{eq:subsampling_p}, as originally proposed in Word2Vec~\cite{Mikolov2013b} models.

\begin{equation}\label{eq:subsampling_p}
    P(\text{discard} \mid i) = (\sqrt{\frac{|U_i|}{\rho}} + 1) \cdot \frac{\rho}{|U_i|}
\end{equation}

\noindent in which $\rho$ corresponds to a hyperparameter used for adjusting the subsampling rate. An increase in the value of $\rho$ increases the probability of removing the interaction, as well as higher values of $|U_i|$.

\subsubsection{Negative sampling}\label{sec:neg_sample}

Predicting scores for every possible user-item pair is computationally demanding, with quadratic growth as new users and items are added. To alleviate this problem, we employed negative sampling to reduce the number of comparisons~\cite{Mikolov2013b}. The weight matrices are updated considering only a small subset $G$ of ``negative'' items (i.e., the ones the user has not interacted with), randomly selected for each ``positive'' item ($i \in I_u$) at each epoch. Therefore, the objective function previously expressed by Equation~\ref{eq:n2v_objetivo} becomes the function presented in Equation~\ref{eq:n2v_objetivo_neg}, which is also maximized.

\begin{equation}\label{eq:n2v_objetivo_neg}
    \frac{1}{|U|}\sum_{u \in U}\Big({\frac{1}{|I_u|}\sum_{i \in I_u}\Big({\text{log } \sigma(u, i)  - \sum_{j \in G} \text{log } \sigma(u, j)}\Big)}\Big)
\end{equation}

The random selection of $G$ follows a probability distribution expressed by Equation~\ref{eq:neg_sampling_distr}, where $z(i)$ represents the number of interactions item $I$ has, that is, $|U_i|$, and $\gamma$ is a hyperparameter that balances the probability of selection between more or less popular items. Although the value for $\gamma$ usually is $0.75$ in NLP applications~\cite{Mikolov2013b}, favoring more popular items to be included in $G$,~\cite{Caselles2018} showed that this might not hold for recommender systems. The authors showed an improvement in the results when using negative values of $\gamma$, which increases the probability of selecting less popular items.

\begin{equation}\label{eq:neg_sampling_distr}
    P(i) = \frac{z(i)^\gamma}{\sum_{j \in I}{z(j)^\gamma}}
\end{equation}

\subsubsection{Regularization}\label{sec:reg_param_n2v}

To avoid over-sampling and improve generalization, we applied L2 regularization (Ridge regression) in both weight matrices $\mathcal{W}$ and $\mathcal{W}'$ of Interact2Vec. The objective function can then be expressed by Equation~\ref{eq:n2v_regularizer}:

\begin{equation}\label{eq:n2v_regularizer}
    \frac{1}{|U|}\sum_{u \in U}\Big({\frac{1}{|I_u|}\sum_{i \in I_u}\Big({\text{log } \sigma(u, i)  - \sum_{j \in G} \text{log } \sigma(u, j)}\Big)}\Big) - \lambda \sum_{i \in I}{||\mathcal{W'}^T_i||^2} - \lambda \sum_{u \in U}{||\mathcal{W}_u||^2}
\end{equation}

\noindent in which $\lambda$ is a hyperparameter that adjusts the impact of the regularization in the content of the embeddings. Higher values for $\lambda$ will give more weight to the magnitude of the embeddings, thus avoiding a large variation of its content values.

\subsection{Recommender algorithms}\label{sec:n2v_recomendacao}

Like Item2Vec and User2Vec, Interact2Vec is not a recommender algorithm itself. Instead, its main goal is to learn vector representations for users and items that other algorithms can use to generate the final recommendation.

While maximizing its objective function, Interact2Vec learns item embeddings that are close in the vector space to the embeddings of the users who have consumed them. This property allows recommender algorithms to use approaches based on similarity to retrieve relevant embeddings or extract knowledge over the formed vector neighborhood.

The following sections present different techniques for using Interact2Vec's embeddings for a top-$N$ recommendation. We propose five different techniques: \textit{(i)} user-item similarity, which compares the users and items embeddings directly; \textit{(ii)} item-item similarity, which ignores the user embeddings and yields the recommendation using only item embeddings; \textit{(iii)} weighted similarity, which combines the prior techniques; \textit{(iv)} embeddings combination, which merges user and item embeddings before calculating item-item similarity; and \textit{(v)} ensemble, which generates the final recommendation using all recommender techniques through a voting system.

\subsubsection{User-item similarity}\label{sec:user_item_sim}

\begin{figure}[!htpb]
    \centering
    \includegraphics[width=0.6\linewidth]{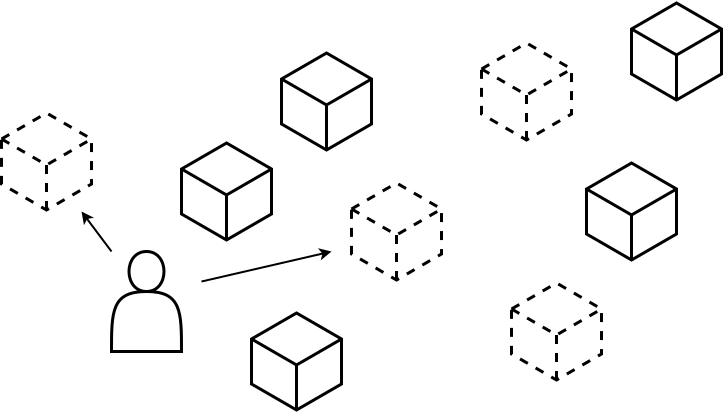}
    \caption{User-item similarity recommendation. Dashed edge cubes represent items the user has not interacted with.}
    \label{fig:user_item_sim}
\end{figure}

Since related users and items are closer in the embeddings vector space, it is possible to generate a top-$N$ ranking for a particular user by simply retrieving the $N$ closest embeddings of items the user has not interacted with, as illustrated in Figure~\ref{fig:user_item_sim}.

Given a user $u$, we calculate the cosine similarity between the user $u$ and every item $i$ in the system. The cosine similarity is defined by Equation~\ref{eq:sim_cosseno}.

\begin{equation}\label{eq:sim_cosseno}
    sim(u,\;i) = \frac{\mathcal{W}_u \cdot \mathcal{W'}^T_i}{||\mathcal{W}_u|| \times ||\mathcal{W'}^T_i||} = \frac{\sum_{l = 1}^{M}{\mathcal{W}_{u_l}\times \mathcal{W'}^T_{i_l}}}{\sqrt{\sum_{l = 1}^{M}{\mathcal{W}_u{_l}}} \times \sqrt{\sum_{l = 1}^{M}{\mathcal{W'}^T_{i_l}}}}
\end{equation}

\noindent in which $\mathcal{W}_u$ represents the user embeddings, and $\mathcal{W'}^T_i$ the item embeddings.

After calculating the similarities, we filter every item $i$ the user has not interacted with, selecting the most similar ones to compose the final recommendation $\text{rec}_u$, as shown in Equation~\ref{eq:top_k_sim}:

\begin{equation}\label{eq:top_k_sim}
    \text{rec}_u = \text{argmax}_N \; (sim(u,\;i) \; \forall \; i \mid i \notin I_u)
\end{equation}

\noindent being $\text{argmax}_N (X)$ an $\text{argmax}$ function applied over $X$ to retrieve $N$ values without replacement.

\subsubsection{Item-item similarity}\label{sec:item_item_sim}

\begin{figure}[!htpb]
    \centering
    \includegraphics[width=0.6\linewidth]{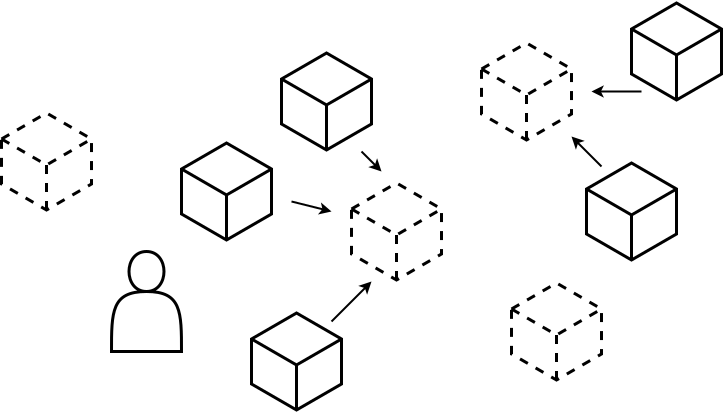}
    \caption{Item-item similarity recommendation. Dashed edge cubes represent items the user has not interacted with.}
    \label{fig:item_item_sim}
\end{figure}

A second possible approach is to ignore the vector representation of users and calculate only the similarities between items, a common strategy used for top-$N$ ranking~\cite{Deshpande2004}. Then, we can compute the recommendation using the similarities between items consumed by the target user and new items, as shown in Figure~\ref{fig:item_item_sim}.

This approach is very similar to the user-item similarity presented in Section~\ref{sec:user_item_sim}, with the final recommendation being generated in the same manner as expressed in Equation~\ref{eq:top_k_sim}. However, the similarity between users and items $sim(u,\;i)$ is calculated differently.

We calculate the cosine similarity $sim(i,\; j)$ through Equation~\ref{eq:sim_cosseno} for every possible pair of items $i$ and $j$ using their embeddings $\mathcal{W'}^T_i$ and $\mathcal{W'}^T_j$. The recommender generates a similarity score for user $u$ and every item $i$ using Equation~\ref{eq:item_item_sim}, filtering the top $N$ items with the higher score.

\begin{equation}\label{eq:item_item_sim}
    sim(u, i) = \frac{\sum_{j \in I_u}{sim(i,\; j)}}{|I_u|}
\end{equation}

\subsubsection{Weighted similarities}

\begin{figure}[!htpb]
    \centering
    \includegraphics[width=0.6\linewidth]{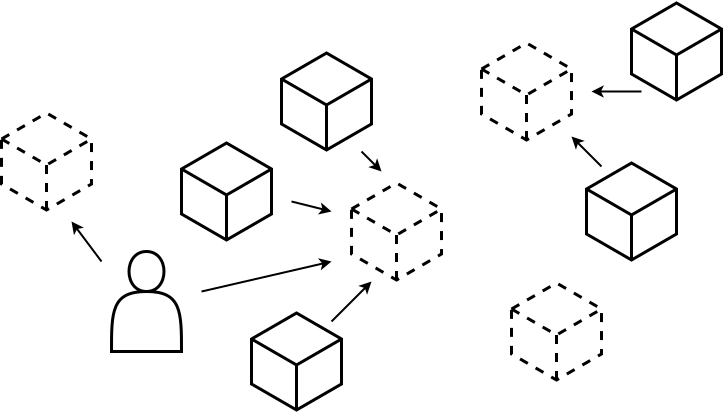}
    \caption{Recommendation by the weighted similarity between user-item and item-item similarities. Dashed edge cubes represent items the user has not interacted with.}
    \label{fig:weighted_sim}
\end{figure}

In many cases, the position of items and users in the vector space can cause divergences between the output of the two aforementioned methods. User--item similarity aims to find the most similar items to a user, while item--item aims to find the most similar items to items previously consumed by that user. To find a better match between these two approaches, we can combine them to build the final recommendation through a weighted vote, as shown in Figure~\ref{fig:weighted_sim}.

In this approach, we can calculate the similarity $sim_w(u,\;i)$ using Equation~\ref{eq:weighted_sim}, i.e., calculating the weighted average between the user-item similarity $sim_u(u,\;i)$, explained in Section~\ref{sec:user_item_sim}, and the item-item similarity $sim_i(u,\;i)$, explained in Section~\ref{sec:item_item_sim}. 

Considering that the performance of user--item or item--item recommendations can vary among different datasets, we propose using $\beta$ and $\mu$ as hyperparameters to regulate the weight that each recommendation method will influence in the final recommendation. Those values can then be optimized in a fine-tuning training phase.

\begin{equation}\label{eq:weighted_sim}
    sim_w(u,\;i) = \frac{\beta \cdot sim_u(u,\;i) + \mu \cdot sim_i(u,\;i)}{\beta + \mu}
\end{equation}

\subsubsection{Combination of the embeddings}

Since user embeddings carry latent information about the items the users have consumed, we can use the average of user embeddings to find points in the vector space that condense a specific consumer behavior. Filtering those users by a particular item can yield a new vector representation of a point in space relevant to that item, which can then be used to enhance the original item embedding.

This strategy consists of recommending through item-item similarity but using new embeddings generated by the combination of user and item embeddings so that a single representation vector carries information about both entities.

This combination is performed by concatenating the average value of multiple user embeddings with the item embedding. Considering $\bar{\mathcal{U}_i}$ as the average embedding of every user $u$ that has consumed $i$, as expressed by Equation~\ref{eq:average_user}, we can represent an item $i$ as $\Big[\;\mathcal{W'}^T_i,\;\bar{\mathcal{U}_i}\Big]$, i.e., the concatenation of the original embedding for $i$ and $\bar{\mathcal{U}_i}$.

\begin{equation}\label{eq:average_user}
    \bar{\mathcal{U}_i} = \frac{\sum_{u \in U_i}\mathcal{W}_u}{|U_i|}
\end{equation}

Additionally, to avoid outliers, it is possible to restrict only the $K$ users nearest to $i$ in the vector space instead of using every user embedding. $K$ then becomes a hyperparameter that can be adjusted during a fine-tuning training step.

\subsubsection{Recommender ensemble}

Finally, since every method mentioned above has a particular strategy to yield the recommendation, it is interesting to compose the final top-$N$ recommendation using a voting mechanism so that the most recommended items among every method are selected.

The ensemble works as follows: for a given user $u$, each method $m$ must yield a top-$L$ ranking $s_m(u, L)$, in which $L \geq N$, i.e., each method must filter a subset of items using its specific recommendation strategy that is greater than the desired size of the final recommendation. 

Being $S$ the set with every top-$L$ recommendation, we can then compute the final recommendation through voting among the selected items in every $s_m(u, L) \in S$. To benefit recommender algorithms that performed better for specific datasets aiming to construct the best recommendation possible, the votes of each method can be weighted according to its performance. Another strategy is to weigh the votes for each item, given its position in the ranking. The final formula for the voting can be seen in Equation~\ref{eq:comite_score}:

\begin{equation}\label{eq:comite_score}
    \text{score}(i,\;u) = \sum_{m = 1}^{|S|}{(f(i,\;u,\;m) \times w_m \times w_r)}
\end{equation}

\noindent in which $f(i,\;u,\;m)$ is a function for checking if $i$ is present in $s_m(u, L)$ (Equation~\ref{eq:comite_f}), $w_m$ is a weight based on the predictive quality of method $m$ and $w_r$ is a weight based on the position of $i$ in the ranking $s_m(u, L)$.

\begin{equation}\label{eq:comite_f}
    f(i,\;u,\;m) = \begin{cases}
        1, & \text{if } i \in s_m(u,\;L) \\ 0, & \text{otherwise}
    \end{cases}
\end{equation}

\subsection{Algorithm efficiency analysis}\label{sec:alg_complexity}

One of the main objectives in designing Interact2Vec is its simplicity and computational efficiency. For this, we proposed a shallow neural network that demands a reduced number of operations for training. To understand this particularity of the model, we present in this section an analysis of the number of operations performed to train the representation, comparing with the complexity of other traditional methods in the area, e.g., $K$-Nearest Neighbors (KNN), Alternating Least Squares (ALS), and Bayesian Personalized Ranking (BPR), as well as other shallow neural embedding models such as Item2Vec and User2Vec, and a state-of-the-art model based on deep learning, RecVAE.

To generalize the comparison and facilitate the calculation of common operations performed by every compared model, we will consider that addition and multiplication are constant operations, having $O(1)$ complexity, and matrix multiplication requires $O(n^3)$ operations. Although it is possible to optimize matrix multiplication using the Strassen algorithm~\cite{Strassen1969}, the new values will not change the comparison between models, thus allowing the cubic solution to be adopted for the explanation. Finally, we will not consider the optimization phase when addressing the efficiency since it depends on the adopted optimization algorithm, and we can use the same optimizer for every model.

To analyze the computational complexity of Interact2Vec, we can split its training procedure into five levels: \textit{(i)} propagation of a single sample; \textit{(ii)} every sample of an item; \textit{(iii)} every item of a user; \textit{(iv)} every user of an epoch; \textit{(v)} every epoch.

The propagation of a single sample, i.e., a user-item pair, consists of a query on a lookup table containing the embeddings, followed by a vector multiplication. Using an indexed table, the query to retrieve the embedding content can be done with $O(1)$ complexity. The vector multiplication is proportional to the embedding size. Therefore, we can represent its complexity as $O(2M)$ or $O(M)$, where $M$ is the dimension for the embeddings.

This operation is performed for a single item $i$ and repeated for every negative item $i \in G$. That way, the network will perform $O(|G| \times M)$ operations. Since the network is adjusted for every item consumed by a user and for every user, we can say that the operation mentioned above is repeated for every interaction $r \in R$. Finally, the network is trained over $C$ epochs. Therefore, we can represent its final complexity as $O(C \times |R| \times |G| \times M)$, i.e., Interact2Vec's complexity is tied to the number of epochs, interactions, negative samples, and the dimensionality of the embedding. In practice, however, the number of interactions is expressively more significant than the number of other factors ($|R| \gg \{M,\;|G|,\;C\}$), so we can say that the number of operations performed by Interact2Vec grows linearly to the number of interactions in the dataset, having $O(|R|)$.

The User2Vec model performs similar operations to Interact2Vec and is also adjusted for each user-item pair in the dataset, having an $O(|R|)$ complexity. Even so, the model has an additional operation: averaging the embeddings of the user and its respective items to be projected in the intermediate layer. However, considering that the average between a set of vectors of size $M$ will have complexity $O(M)$, this operation has no real impact on the final complexity (although, in practice, this may cause Interact2Vec to achieve a slightly better performance).

The Item2Vec model works similarly to Interact2Vec but has a significant difference. Since it is trained by receiving an item consumed by a user and predicting the other items consumed by them, the model performs $|I_u|-1$ propagation steps for each interaction between a user $u$ and an item $i$. As the rest of Item2Vec's operations are similar to Interact2Vec, the computational complexity of the model is approximately $O(C \times |U| \times (|I_u|-1) \times |G| \times M) $, thus being conditioned to the number of items consumed by each user. In the worst-case scenario where each user consumes all items in the catalog, there is $O(C \times |R| \times (|R|-1) \times |G| \times M) \approx O(|R|^2)$, having quadratic performance. In practice, the number of items consumed by each user will be smaller than those in the catalog, guaranteeing a complexity of approximately $O(|R|)$. However, Interact2Vec will always perform better than Item2Vec, as it performs fewer operations, and its advantage will only increase in scenarios where users interact with many items.

Regarding matrix factorization techniques for implicit recommendation, most operations are performed when learning the user and item latent factors. The implicit ALS algorithm~\citep{Hu2008} has a complexity of $O(|U| \times |I|)$, making its application unfeasible in real scenarios. In this way, the authors proposed techniques to optimize it, reaching complexity $O(M^2 \times |R| + M^3)$ for each of the $C$ iterations, where $M$ corresponds to the number of latent factors, thus presenting a complexity of $O(C \times M^2 \times |R| + C \times M^3)$. Similar to Interact2Vec, the number of operations performed by ALS grows linearly according to the number of interactions since $|R| \gg \{M,\;C\}$. Still, in scenarios where the number of ALS latent factors is greater than the number of negative samples selected by Interact2Vec (i.e., $M > |G|$), the neural model will perform a slightly lower number of operations, thus being able to present greater efficiency than the matrix factorization method.

BPR~\cite{Rendle2009}, another popular matrix factorization method for implicit feedback, performs a user-item relationship prediction step for each tuple $(\text{user}, \; \\ \text {item}_1, \; \text{item}_2)$. This prediction can be computed in different ways, with the most efficient being by calculating the vector product between the latent factors of the representation vectors, which has complexity $O(M)$. If performed for all possible sets of user and item pair, the operation would be performed $|R| \times |I|$ times over $C$ iterations, thus obtaining a computational complexity of $O(C \times |R| \times I \times M)$. In this scenario, BPR would be extremely more expensive than Interact2Vec since $|I| \gg |G|$. However, the authors of the method proposed the use of bootstrap resampling to speed up the execution, selecting a reduced and random number of samples and terminating the execution of the method in case of convergence, which, according to experiments, causes the number of method operations grows linearly at $|R|$, obtaining behavior similar to Interact2Vec, that is, $O(|R|)$.

When analyzing a deep learning approach, the time complexity heavily depends on the model's architecture, such as the width of the neural network and the number of neurons. This study compared Interact2Vec with RecVAE~\cite{Shenbin2019}, a state-of-the-art variational autoencoder for implicit recommendation. The model consists of two fully connected neural networks, the encoder and the decoder. Disregarding parallelization, the time complexity of training a traditional multilayer perceptron is $O(S \times L  \times H^2 \times C)$, in which $S$ corresponds to the number of samples, $L$ the number of layers, $H$ the number of neurons on the hidden layers and $C$ the number of epochs. For RecVAE, the number of samples is related to the number of users $|U|$; the encoder is a deep network with a variable size, while the decoder consists of a shallow network of only one layer. Considering the same number of neurons as Interact2Vec's embedding size, we can describe the time complexity of RecVAE as $O(|U| \times L \times M^2 \times C + |U| \times M^2 \times C) \approx O(2|U|M^2L$). As opposed to Interact2Vec, the final complexity of RecVAE is influenced by the number of users instead of the number of interactions and scales quadratically according to the number of neurons on the hidden layers. This can result in better performance in scenarios where a single user consumes multiple items, but it may become unfeasible when the architecture grows in width and number of neurons.

Finally, KNN is the only compared method that does not need to generate a vector representation of users or items to generate the recommendation. Still, since it uses high-dimensional vectors according to the number of items or users in the catalog, its recommendation step is performed over matrices of shape $|U| \times |I|$. In contrast, matrix factorization or neural embedding models use reduced-dimensional matrices of shape $|U| \times M$ or $|I| \times M$. Given that $M \ll |U| \text{ ou } |I|$, a KNN recommendation over the full interaction matrix will always be less computationally efficient than any embedding model.

We can then conclude that Interact2Vec is more efficient -- or at least similar -- to the baseline recommenders, being a good option for generating user and item embeddings with computational efficiency, which can benefit applying the model in scenarios with limited resources.

\subsection{Limitations}

Interact2Vec is a distributed vector representation model that can learn item and user embeddings simultaneously and efficiently. Although it has advantages over other models, it has certain disadvantages and limitations that should be carefully observed when applying it in real scenarios.

The model has several hyperparameters that need to be fine-tuned: the size $M$ of the embeddings, the subsampling rate of frequent items $\rho$, the number of negative samples $|G|$, the power of the probability distribution of negative samples $\gamma$ and the regularization parameter $\lambda$, in addition to other general characteristics of neural networks, such as the number of iterations, the optimizer algorithm, the learning rate and the batch size.

Having many hyperparameters to be adjusted is not exclusive to Interact2Vec. Item2Vec, User2Vec, and several other embedding-based or neural network-based methods also share this property. However, this can be seen as a pitfall compared to more straightforward methods, such as KNN. On the other hand, the high number of parameters gives the method greater adaptability to different scenarios.

Another essential factor to be observed is the concept drift problem, i.e., when the underlying patterns in the data evolve, causing a previously trained model to become less accurate as it no longer reflects the current reality. Interact2Vec is susceptible to this issue because its embedding learning phase relies on user-item interactions, and user-based methods typically require more frequent updates to maintain performance~\cite{Sarwar2001,Grbovic2015}. This is because user preferences tend to change more rapidly than item characteristics, an advantage of item-based methods like Item2Vec and KNN, which are less affected by user dynamics. Furthermore, Interact2Vec is not designed to handle the cold-start problem effectively, as it cannot incrementally learn embeddings for new users or items added to the dataset.

The primary strategy to mitigate these issues is to employ the embeddings generated by Interact2Vec in item-based recommender algorithms, such as the one explained in Section~\ref{sec:item_item_sim}. Interact2Vec is also an efficient model that can be frequently updated, effectively addressing the concept drift problem. Finally, although user-based models may become outdated more rapidly, generating user embeddings can be valuable for other applications within the area of recommender systems, such as user clustering.

\section{Experiments and results}\label{resultados}

This section describes the experimental protocol for comparing Interact2Vec with other embedding-based models. First, we have conducted a top-$N$ ranking scenario to address the quality of the method in generating the recommendation itself. Next, we have performed an intrinsic analysis of the content of the embeddings, using similarity tables and automatic feature prediction to address how well Interact2Vec learns the embeddings in retaining context information. We present both results and the experimental protocol in the following sections. The source code used to run the experiments, as well as the implementation of the models, is publicly available at \url{https://github.com/ReisPires/Interact2Vec}.

\subsection{Datasets and preprocessing}

We conducted the experiments over the datasets in Table~\ref{tab:dados}. All datasets are widely used in literature, publicly accessible, and consist of different application domains. Columns $|U|$, $|I|$, and $|R|$ contain, respectively, the number of users, items, and interactions. Column $S$ shows the sparsity rate (i.e., $S = 1 - \frac{|R|}{|U| \times |I|}$). Finally, column ``Type'' indicates if the interactions of the dataset are explicit (E), implicit (I), or a combination of both (E+I).

\begin{table}[htpb]
\centering
\begin{tabular}{@{}lccccc@{}}
\toprule
\textbf{Dataset} & $\boldsymbol{|U|}$ & $\boldsymbol{|I|}$ & $\boldsymbol{|R|}$ & $\boldsymbol{S}$ & \textbf{Type} \\ \midrule
Anime\footnotemark[1] & 37,128 & 10,697 & 1,476,495 & 99.63\% & E+I  \\ 
BestBuy\footnotemark[2] & 1,268,702 & 69,858 & 1,862,782 & 99.99\% & I  \\ 
BookCrossing\footnotemark[3] & 59,517 & 246,724 & 716,109 & 99.99\% & E+I   \\ 
CiaoDVD\footnotemark[4] & 17,615 & 16,121 & 72,345 & 99.97\% & E \\ 
Delicious\footnotemark[3] & 1,867 & 69,223 & 104,799 & 99.92\% & I \\ 
Filmtrust\footnotemark[4] & 1,508 & 2,071 & 35,494 & 98.86\% & E   \\ 
Last.FM\footnotemark[3] & 1,892 & 17,632 & 92,834 & 99.72\% & I \\ 
MovieLens\footnotemark[3] & 162,541 & 59,047 & 25,000,095 & 99.74\% & E  \\ 
NetflixPrize\footnotemark[5] & 480,189 & 17,770 & 100,480,507 & 98.82\% & E  \\ 
RetailRocket\footnotemark[6] & 11,719 & 12,025 & 21,270 & 99.98\% & I  \\ \bottomrule
\end{tabular}
\caption{Datasets used in the experiments.}
\label{tab:dados}
\end{table}

\footnotetext[1]{Anime Recommendations dataset. Available at: \url{https://www.kaggle.com/datasets/CooperUnion/anime-recommendations-database}. Accessed on \today.}
\footnotetext[2]{Data Mining Hackathon on BIG DATA (7GB). Available at: \url{www.kaggle.com/c/acm-sf-chapter-hackathon-big}. Accessed on \today.}
\footnotetext[3]{GroupLens - Datasets. Available at: \url{www.grouplens.org/datasets/}. Accessed on \today.} 
\footnotetext[4]{CiaoDVD dataset and Filmtrust. Available at: \url{www.github.com/caserec/Datasets-for-Recommender-Systems}. Accessed on \today.}
\footnotetext[5]{Netflix Prize data. Available at: \url{www.kaggle.com/netflix-inc/netflix-prize-data}. Accessed on \today.}
\footnotetext[6]{Retailrocket recommender system dataset. Available at \url{www.kaggle.com/retailrocket/ecommerce-dataset}. Accessed on \today.}

During preprocessing, we removed all duplicated interactions, keeping only one occurrence. We removed every inconsistent interaction, i.e., the same user-item pair with different ratings. For datasets with different types of interaction, we used only the one that best indicates genuine user interest, i.e., ``listen'' for Last.FM and ``buy'' for RetailRocket. This choice helps filter out weaker or more ambiguous signals, such as ``tagged'' and ``view'', which may not reliably reflect user preference. While this simplification may exclude some potentially useful information, it focuses the training and evaluation on high-confidence interactions. Moreover, because all recommendation models are trained on the same filtered data, and none of them explicitly account for interaction types, this decision is not expected to bias the comparative results of the experiments. In addition, we considered every interaction as positive for datasets containing explicit feedback, no matter the associated rating.

\subsection{Experimental protocol}

We split the datasets into train, validation, and test sets, following an 8:1:1 rate, and trained the models using grid-search cross-validation. Due to computational power limitations, we used only a randomly selected subset of interactions for MovieLens and NetflixPrize datasets, 10\% and 5\% during the hyperparameter adjustment phase and 100\% and 25\% during the final experiment, respectively. Moreover, we excluded items consumed by a single user and users who interacted with only one item due to the limitations of specific benchmark methods. These require training with users with two or more interactions with the system. Since we did not want to address the cold-start problem, we removed instances of unknown users and items from both the validation and test datasets.

Interact2Vec was compared with six other algorithms: one based on neighborhood, $K$-Nearest Neighbors (KNN)~\cite{Sarwar2002}; two based on matrix factorization, implicit Alternating Least Squares (ALS)~\cite{Hu2008} and Bayesian Personalized Ranking (BPR)~\cite{Rendle2009}; two based on neural embeddings, Item2Vec (IT2V)~\cite{Barkan2016} and User2Vec (US2V)~\cite{Grbovic2015}; and one state-of-the-art variational autoencoder, RecVAE~\cite{Shenbin2019}. It is essential to highlight that the performance comparison of Interact2Vec should be primarily performed between neural embedding-based methods. The performance of traditional and deep learning approaches is just necessary to position the performance of Interact2Vec against other well-known strategies.

All algorithms were implemented in \texttt{Python3}. For KNN we used the implementation of \textit{turicreate library}. Both ALS and BPR were implemented using \textit{implicit library}, with latent factors $f$ ranging in $\{50, 100, 200\}$, regularization factor $\lambda$ in $\{10^{-6}, 10^{-4}, 10^{-2}\}$ and learning rate $\alpha$ in $\{0.0025, 0.025, 0.25\}$. IT2V and US2V were implemented using \textit{gensim library}, with $C$ ranging in $\{50, 100, 150\}$, $\gamma$ in $\{-1.0, -0.5, 0.5, 1.0\}$, and $\rho$ in $\{10^{-5}, 10^{-4}, 10^{-3}\}$, while the remaining hyperparameters followed insights proposed in~\cite{Caselles2018}, with fixed optimal values for the parameters shown in Table~\ref{tab:parametrosBaseOtimizacaoEmb}. Lastly, RecVAE (RVAE) was tested with its original implementation\footnote{RecVAE source code. Available at: \url{https://github.com/ilya-shenbin/RecVAE}. Accessed on \today.}, with $\gamma$ ranging in $\{0.0035, 0.005, 0.01\}$ and the remaining hyperparameters fixed to the values recommended by its authors~\cite{Shenbin2019}.

\begin{table}[htpb!]
    \centering
    \begin{tabular}{@{}lcc@{}}
        \toprule
        \textbf{Hyperparameter} & \textbf{Notation} & \textbf{Default value} \\ \midrule
        Learning rate & $\alpha$ & 0.25 \\ 
        Embeddings size & $M$ & 100 \\ 
        Number of negative samples & $|G|$ & 5 \\ \bottomrule
    \end{tabular}
    \caption{Default values used in the literature for embedding-based neural models for recommendation~\cite{Caselles2018}}
    \label{tab:parametrosBaseOtimizacaoEmb}
\end{table}

In the context of Interact2Vec (IN2V), given the computational impracticality of exhaustively testing all parameter combinations, we undertook a comprehensive study to evaluate the impact of parameter selection, which can be seen in~\ref{apx:in2v_parameters}. Our investigation revealed that specific parameter values significantly influence the model's performance. Notably, optimal results were observed for the learning rate ($\alpha$), the number of epochs ($C$), the subsampling rate for frequent items ($\rho$), and the regularization factor ($\lambda$), which can be seen in Table~\ref{tab:parametrosBaseOtimizacaoIN2V} and were fixed throughout the experiments. We then tested different values for the size of the embeddings ($M$), which ranged in the set $\{50, 100, 150\}$, the number of negative samples ($G$) in $\{5, 10, 15\}$, and the negative exponent for negative sampling ($\gamma$) in $\{-1.0, -0.5, 0.5, 1.0\}$.

\begin{table}[htpb!]
    \centering
    \label{tab:parametrosBaseOtimizacaoIN2V}
    \begin{tabular}{@{}lcc@{}}
        \toprule
        \textbf{Hyperparameter} & \textbf{Notation} & \textbf{Default value} \\ \midrule
        Learning rate & $\alpha$ & 0.25 \\
        Number of epochs & $C$ & 50 \\
        Subsampling rate for frequent items & $\rho$ & $10^{-6}$ \\
        Regularization factor & $\lambda$ & 0.1 \\
        \bottomrule
    \end{tabular}
    \caption{Ideal values for Interact2Vec's hyperparameters}
\end{table}

In addition to fine-tuning parameters during the embedding learning phase, adjustments were made to hyperparameters related to the recommender algorithms responsible for consuming the embeddings and generating final recommendations. For IT2V and US2V, item-item and user-item similarity were employed, respectively, as proposed by the authors in their original work. In the case of IN2V, a comprehensive assessment was conducted across all algorithms listed in Section~\ref{sec:n2v_recomendacao}. For the weighted similarities, we explored different values for $\beta$ and $\mu$; for the combination of the embeddings, we experimented with multiple values for $K$; and for the recommender ensemble, we tuned the neighborhood-size $L$ and considered using weighted votes according to the quality of the method and the position of the item. All the tested values can be seen in Table~\ref{tab:parametrosMetodosRecIN2V}. The selection of the optimal recommender algorithm for each dataset was determined through a grid search, considering the performance on the validation set. Results exhibited significant variability across diverse datasets without a consensus on the best recommender algorithm. Nevertheless, preliminary findings suggest the potential superiority of recommendations using weighted similarities, a trend we intend to explore further in future work.

\begin{table}[htpb]
    \centering

    \begin{tabular}{@{}ll@{}}
        \toprule
        \textbf{Recommender algorithm} & \textbf{Tested values} \\ \midrule
        \multirow{2}{*}{Weighted similarities} & $\beta = \{0.1,\;0.25,\;0.5,\;0.75,\;0.9\}$ \\
        & $\mu = \{0.1,\;0.25,\;0.5,\;0.75,\;0.9\}$ \\ \midrule
        \multirow{1}{*}{Combination of embeddings} & 
        $K = \{1, 5, 10, 15, |U_i|\}$ \\ \midrule
        \multirow{3}{*}{Recommender ensemble} & use of $w_m = \{\text{yes},\;\text{no}\}$ \\
        & use of $w_r = \{\text{yes},\;\text{no}\}$ \\
        & $L = \{15,\;30,\;45\}$ \\
        \bottomrule
    \end{tabular}
    \caption{Tested values during parameter tuning of the recommender algorithms consuming Interact2Vec's embeddings}
    \label{tab:parametrosMetodosRecIN2V}
\end{table}

\subsection{Top-$N$ ranking task}

To evaluate the ability of the algorithms to generate recommendations, we calculated the F1-score and NDCG in a top-$N$ recommendation scenario, with $N$ ranging between $1 \text{ and } 20$. 

The \textbf{F1-score} consists of a harmonic mean between precision and recall, with the two metrics describing how well the recommendations yielded by the algorithms are when compared to what the user consumed. Precision represents how many recommended items were correct hits, while recall represents how many consumed items were included in the recommendation. Considering that the former tends to decrease for greater values of $N$ since more items are being recommended and the latter tends to increase since more consumed items can be included in the recommendation, the F1-score can balance both metrics. \textbf{Normalized Discounted Cumulative Gain (NDCG)}, in addition to evaluating hits like F1, also assesses ranking, favoring algorithms in which correct recommendations were present in the top spots of the recommendation list. This is done by weighting every hit according to the inverse logarithm of its position in the recommendation.

The models showed consistent behavior across various values of $N$, with minor variations in their relative positions. Given this stability, our analysis concentrated on a top-$15$ scenario, a frequently adopted threshold in the literature. This choice allows for a more focused examination while aligning with established practices in the field. Additionally, the recommenders behaved similarly in both metrics, i.e., methods that achieved good results for F1 tended to achieve good results for NDCG. Results are shown in the grayscale Tables~\ref{tab:f1_at_15} and~\ref{tab:ndcg_at_15}, for the F1@15 and NDCG@15, respectively. The darker the cell, the better the result for that specific dataset and metric, with the best result highlighted in \textbf{bold}.

\begin{table}[!htb]
\centering
\begin{tabular}{r|ccc|ccc|c}
\toprule
\multirow{2}{*}{\textbf{Dataset}} & \multicolumn{3}{c|}{\textbf{Traditional}} & \multicolumn{3}{c|}{\textbf{Embedding-based}} & \multicolumn{1}{c}{\textbf{SOTA}} \\
& \textbf{KNN} & \textbf{ALS} & \textbf{BPR} & \textbf{IT2V} & \textbf{US2V} & \textbf{IN2V} & \textbf{RVAE} \\ \midrule
\textbf{Anime} & \cellcolor[gray]{0.572}0.133 & \cellcolor[gray]{0.568}0.134 & \cellcolor[gray]{0.701}0.090 & \cellcolor[gray]{0.707}0.088 & \cellcolor[gray]{0.95}0.007 & \cellcolor[gray]{0.745}0.075 & \cellcolor[gray]{0.5}\textbf{0.156} \\
\textbf{BestBuy} & \cellcolor[gray]{0.539}0.034 & \cellcolor[gray]{0.822}0.014 & \cellcolor[gray]{0.789}0.016 & \cellcolor[gray]{0.777}0.017 & \cellcolor[gray]{0.95}0.005 & \cellcolor[gray]{0.838}0.013 & \cellcolor[gray]{0.5}\textbf{0.036} \\
\textbf{BookCrossing} & \cellcolor[gray]{0.622}0.007 & \cellcolor[gray]{0.5}\textbf{0.010} & \cellcolor[gray]{0.717}0.006 & \cellcolor[gray]{0.601}0.008 & \cellcolor[gray]{0.95}0.001 & \cellcolor[gray]{0.786}0.004 & \cellcolor[gray]{0.553}0.009 \\
\textbf{CiaoDVD} & \cellcolor[gray]{0.764}0.011 & \cellcolor[gray]{0.665}0.015 & \cellcolor[gray]{0.743}0.012 & \cellcolor[gray]{0.74}0.012 & \cellcolor[gray]{0.95}0.003 & \cellcolor[gray]{0.571}0.019 & \cellcolor[gray]{0.5}\textbf{0.022} \\
\textbf{Delicious} & \cellcolor[gray]{0.919}0.060 & \cellcolor[gray]{0.943}0.023 & \cellcolor[gray]{0.946}0.020 & \cellcolor[gray]{0.932}0.039 & \cellcolor[gray]{0.95}0.013 & \cellcolor[gray]{0.922}0.054 & \cellcolor[gray]{0.5}\textbf{0.691} \\
\textbf{Filmtrust} & \cellcolor[gray]{0.533}0.268 & \cellcolor[gray]{0.95}0.110 & \cellcolor[gray]{0.604}0.241 & \cellcolor[gray]{0.507}0.278 & \cellcolor[gray]{0.906}0.127 & \cellcolor[gray]{0.528}0.270 & \cellcolor[gray]{0.5}\textbf{0.280} \\
\textbf{Last.FM} & \cellcolor[gray]{0.5}\textbf{0.120} & \cellcolor[gray]{0.614}0.094 & \cellcolor[gray]{0.68}0.078 & \cellcolor[gray]{0.567}0.105 & \cellcolor[gray]{0.95}0.016 & \cellcolor[gray]{0.57}0.104 & \cellcolor[gray]{0.518}0.116 \\
\textbf{MovieLens} & \cellcolor[gray]{0.598}0.160 & \cellcolor[gray]{0.564}0.175 & \cellcolor[gray]{0.709}0.110 & \cellcolor[gray]{0.735}0.098 & \cellcolor[gray]{0.95}0.000 & \cellcolor[gray]{0.82}0.059 & \cellcolor[gray]{0.5}\textbf{0.204} \\
\textbf{NetflixPrize} & \cellcolor[gray]{0.524}0.057 & \cellcolor[gray]{0.636}0.042 & \cellcolor[gray]{0.716}0.031 & \cellcolor[gray]{0.546}0.054 & \cellcolor[gray]{0.95}0.001 & \cellcolor[gray]{0.734}0.029 & \cellcolor[gray]{0.5}\textbf{0.060} \\
\textbf{RetailRocket} & \cellcolor[gray]{0.609}0.022 & \cellcolor[gray]{0.607}0.022 & \cellcolor[gray]{0.95}0.002 & \cellcolor[gray]{0.676}0.018 & \cellcolor[gray]{0.88}0.006 & \cellcolor[gray]{0.701}0.017 & \cellcolor[gray]{0.5}\textbf{0.029} \\ \bottomrule
\end{tabular}
\caption{F1@15 achieved by each algorithm in each dataset.}
\label{tab:f1_at_15}
\end{table}

\begin{table}[!htb]
\centering
\begin{tabular}{r|ccc|ccc|c}
\toprule
\multirow{2}{*}{\textbf{Dataset}} & \multicolumn{3}{c|}{\textbf{Traditional}} & \multicolumn{3}{c|}{\textbf{Embedding-based}} & \multicolumn{1}{c}{\textbf{SOTA}} \\
& \textbf{KNN} & \textbf{ALS} & \textbf{BPR} & \textbf{IT2V} & \textbf{US2V} & \textbf{IN2V} & \textbf{RVAE} \\ \midrule
\textbf{Anime} & \cellcolor[gray]{0.569}0.212 & \cellcolor[gray]{0.523}0.237 & \cellcolor[gray]{0.642}0.174 & \cellcolor[gray]{0.728}0.128 & \cellcolor[gray]{0.95}0.008 & \cellcolor[gray]{0.729}0.127 & \cellcolor[gray]{0.5}\textbf{0.250} \\
\textbf{BestBuy} & \cellcolor[gray]{0.525}0.142 & \cellcolor[gray]{0.79}0.063 & \cellcolor[gray]{0.752}0.075 & \cellcolor[gray]{0.816}0.056 & \cellcolor[gray]{0.95}0.016 & \cellcolor[gray]{0.857}0.044 & \cellcolor[gray]{0.5}\textbf{0.149} \\
\textbf{BookCrossing} & \cellcolor[gray]{0.572}0.015 & \cellcolor[gray]{0.508}0.017 & \cellcolor[gray]{0.717}0.010 & \cellcolor[gray]{0.625}0.013 & \cellcolor[gray]{0.95}0.001 & \cellcolor[gray]{0.778}0.008 & \cellcolor[gray]{0.5}\textbf{0.018} \\
\textbf{CiaoDVD} & \cellcolor[gray]{0.729}0.031 & \cellcolor[gray]{0.639}0.041 & \cellcolor[gray]{0.74}0.030 & \cellcolor[gray]{0.756}0.028 & \cellcolor[gray]{0.95}0.006 & \cellcolor[gray]{0.595}0.046 & \cellcolor[gray]{0.5}\textbf{0.056} \\
\textbf{Delicious} & \cellcolor[gray]{0.589}0.159 & \cellcolor[gray]{0.867}0.055 & \cellcolor[gray]{0.888}0.047 & \cellcolor[gray]{0.755}0.097 & \cellcolor[gray]{0.95}0.024 & \cellcolor[gray]{0.583}0.162 & \cellcolor[gray]{0.5}\textbf{0.193} \\
\textbf{Filmtrust} & \cellcolor[gray]{0.561}0.559 & \cellcolor[gray]{0.903}0.238 & \cellcolor[gray]{0.563}0.557 & \cellcolor[gray]{0.547}0.572 & \cellcolor[gray]{0.95}0.195 & \cellcolor[gray]{0.553}0.566 & \cellcolor[gray]{0.5}\textbf{0.616} \\
\textbf{Last.FM} & \cellcolor[gray]{0.5}\textbf{0.222} & \cellcolor[gray]{0.58}0.186 & \cellcolor[gray]{0.64}0.160 & \cellcolor[gray]{0.573}0.189 & \cellcolor[gray]{0.95}0.023 & \cellcolor[gray]{0.608}0.174 & \cellcolor[gray]{0.537}0.205 \\
\textbf{MovieLens} & \cellcolor[gray]{0.648}0.216 & \cellcolor[gray]{0.568}0.273 & \cellcolor[gray]{0.714}0.168 & \cellcolor[gray]{0.781}0.121 & \cellcolor[gray]{0.95}0.000 & \cellcolor[gray]{0.855}0.068 & \cellcolor[gray]{0.5}\textbf{0.321} \\
\textbf{NetflixPrize} & \cellcolor[gray]{0.5}\textbf{0.069} & \cellcolor[gray]{0.575}0.058 & \cellcolor[gray]{0.67}0.043 & \cellcolor[gray]{0.53}0.064 & \cellcolor[gray]{0.95}0.001 & \cellcolor[gray]{0.722}0.036 & \cellcolor[gray]{0.601}0.054 \\
\textbf{RetailRocket} & \cellcolor[gray]{0.594}0.110 & \cellcolor[gray]{0.551}0.123 & \cellcolor[gray]{0.95}0.002 & \cellcolor[gray]{0.684}0.083 & \cellcolor[gray]{0.914}0.013 & \cellcolor[gray]{0.739}0.066 & \cellcolor[gray]{0.5}\textbf{0.138} \\ \bottomrule
\end{tabular}
\caption{NDCG@15 achieved by each algorithm in each dataset.}
\label{tab:ndcg_at_15}
\end{table}

Interact2Vec (IN2V) presented results comparable to the other methods, achieving the second or third-best performance in 30\% of the datasets. When compared to other neural embedding-based algorithms, Interact2Vec achieved promising results. It was superior to User2Vec (US2V) in every dataset and metric and presented a similar performance to Item2Vec.

Compared with the remaining methods based on matrix factorization or deep learning, the results achieved by Interact2Vec were mixed. RecVAE and KNN were superior in most datasets, which is reasonable considering that the former is a deep-learning model that can learn complex and non-linear patterns, and the latter is a recommender algorithm that does not employ dimensionality reduction, retaining the full high-dimensional structure of the data and avoiding the potential loss of information. However, both RecVAE and KNN have the drawback of being computationally expensive. In addition, as the volume of data increases, KNN becomes progressively inefficient because it requires distance calculations between large-dimensional vector representations. This computational inefficiency is a notable limitation in large-scale and real-world scenarios, similar to or even worse than the datasets used in our experiments. In contrast, the other models we evaluated incorporate dimensionality reduction techniques that mitigate this inefficiency, making them more scalable and suitable for handling large datasets.

When compared with matrix factorization algorithms, ALS was a slightly superior method than Interact2Vec, achieving the third-best NDCG in most datasets. However, the method performed poorly in certain cases, such as Delicious and Filmtrust. BPR, on the other hand, was worse than Interact2Vec in 50\% of the experiments, with some great discrepancies as seen in the RetailRocket dataset.

When analyzing the scenarios in which the proposed model performed better or worse, we can find patterns regarding its performance and the size of the datasets. Anime, BestBuy, BookCrossing, MovieLens, and NetflixPrize, in which IN2V presented suboptimal results, are datasets with the most significant number of interactions, as seen in Table 2. This indicates that, although its computational performance is superior to other methods, as described in Section 3.5 and in the following experiments, the neural model tends to achieve inferior results in datasets with a massive volume of data. While the exact reason for this behavior remains unclear, one possible explanation is that increasing the number of interactions affects the negative sampling step, making it more frequent for an item to be drawn as a negative interaction, which can hinder the model's ability to learn robust representations. Further investigation is needed to understand this phenomenon fully, and we consider it a key direction for future work.

When we observe the results for datasets with fewer interactions, which can even be considered more challenging since less knowledge is available, IN2V tends to perform better. Of the smaller datasets, the ones that IN2V presented the best or second-best results, i.e., CiaoDVD, Delicious, and FilmTrust, have the similarity of having a greater distribution in the average number of items consumed by users, as shown in the boxplots presented in Figure 6. Although Last.FM and RetailRocket datasets have less data, they have a much more flattened distribution of average interactions per user. Considering that Interact2Vec learns the vector representation only by consuming user--item pairs, it is expected that the model would behave better when users tend to consume more items on average.

\begin{figure}[htb]
	\centering
	\includegraphics[scale=0.35]{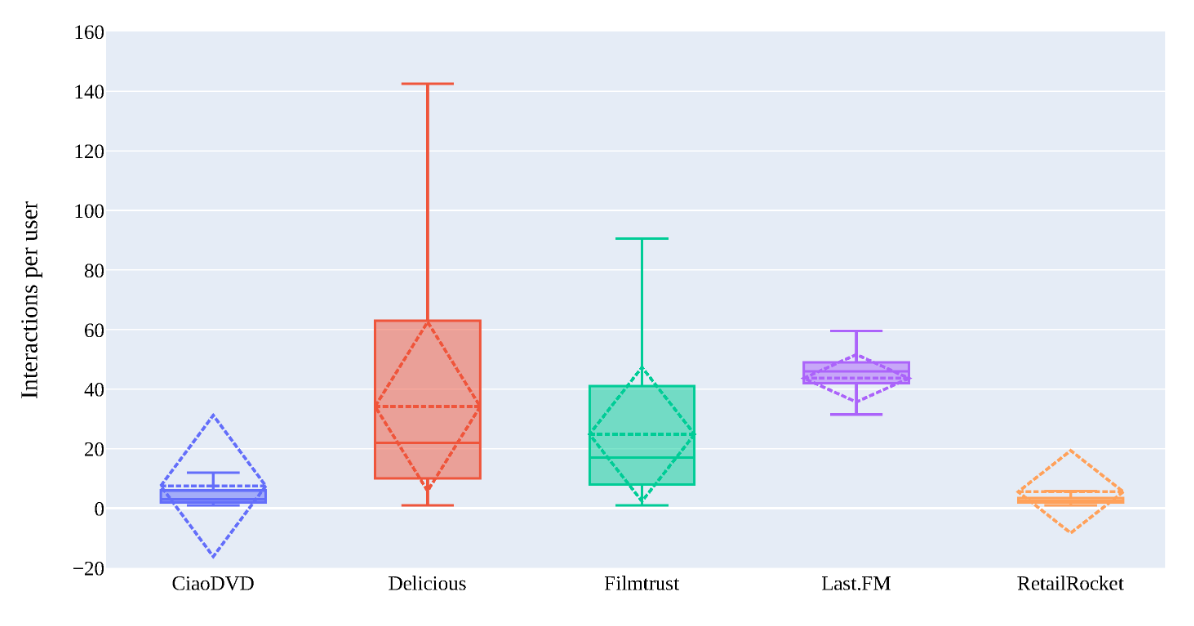}
	\caption{Distribution of the number of items consumed by each user in datasets with fewer interactions.}
	\label{fig:dados_boxplot}
\end{figure}

\begin{figure}[!htpb]
	\centering
	\includegraphics[scale=0.45]{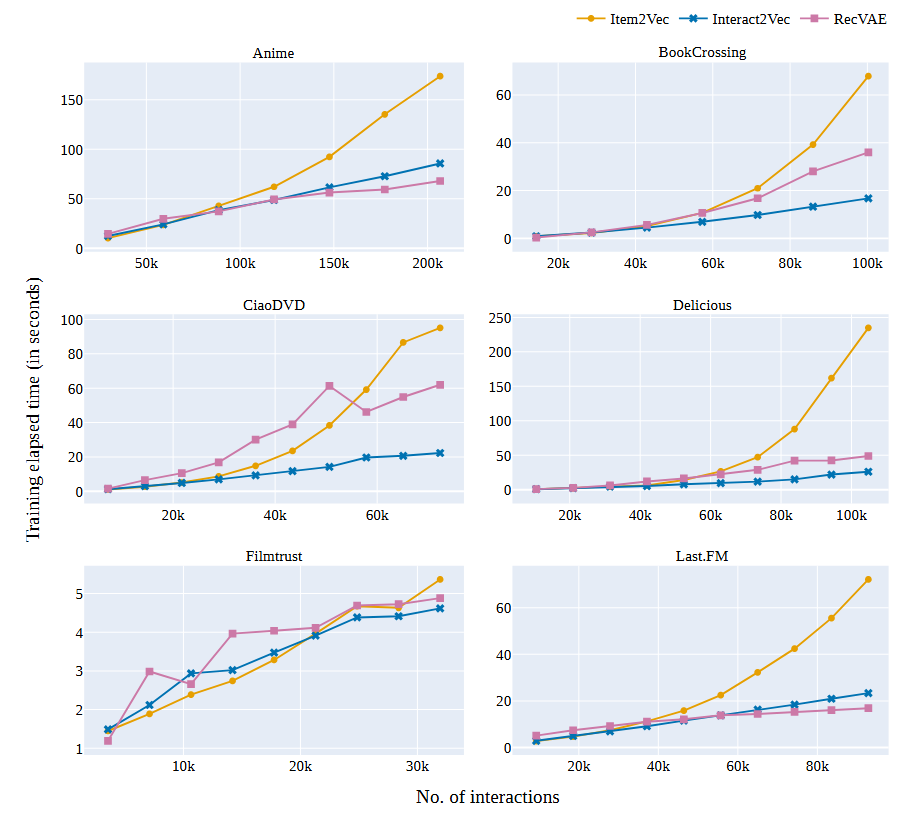}
	\caption{Comparison of the elapsed time for training Item2Vec, RecVAE and Interact2Vec when increasing the number of interactions in different datasets.}
	\label{fig:itm2v_vs_int2v_elapsed_time}
\end{figure}

Given that computational efficiency is one of the main goals when proposing Interact2Vec, we conducted a practical evaluation to compare it with Item2Vec since they are neural embedding-based models that achieved similar results in the top-$N$ ranking task, and RecVAE, since it is a state-of-the-art model. For this, we have implemented both models using the same optimization techniques and fixed values for similar hyperparameters, thus creating a comparable environment. We trained the models in different datasets using incremental sets of samples. For each combination of model, dataset, and sampling, we conducted three independent executions of the training phase and stored the elapsed time. The average elapsed time of each execution is depicted in Figure~\ref{fig:itm2v_vs_int2v_elapsed_time}, with each plot corresponding to a different dataset. Results show that the greater the number of interactions, the more discrepant the superiority of Interact2Vec is when compared to Item2Vec. In specific scenarios, the proposed model could train the embeddings more than 800\% faster than Item2Vec while presenting similar NDCG, a very high performance considering the gains in computational complexity described in Section~\ref{sec:alg_complexity}. Compared to RecVAE, Interact2Vec was more efficient in 66\% of the experiments, with a significant superiority for BookCrossing and CiaoDVD datasets. As stated in Section~\ref{sec:alg_complexity}, this efficiency gap reduces in cases where the average number of items consumed by users increases, as seen in Anime and Last.FM, in which the autoencoder model surpassed Interact2Vec in the last iterations. Even so, on average, Interact2Vec presented a training time of 53.41\% faster than RecVAE. The comparisons between the elapsed time for training the models and the NDCG@15 are presented in Table~\ref{tab:itm2v_vs_int2v}.

\begin{table}[!htb]
\small
\centering
\begin{tabular}{@{}r|rr|rr@{}}
\toprule
\multirow{2}{*}{\textbf{Dataset}} & \multicolumn{2}{c|}{\textbf{Item2Vec}} & \multicolumn{2}{c}{\textbf{RecVAE}} \\
 &  \multicolumn{1}{c}{\textbf{NDCG@15}} &  \multicolumn{1}{c|}{\textbf{Training time}} &  \textbf{NDCG@15} &  \textbf{Training time} \\ \midrule
\textbf{Anime       } & -0.63\%  & +110.26\% & -49.22\%  & -9.86\% \\ 
\textbf{BookCrossing} & -41.98\%  & +113.79\% & -56.82\%  & +65.52\% \\ 
\textbf{CiaoDVD     } & +64.98\%  & +361.90\% & -18.83\%  & +190.48\% \\ 
\textbf{Delicious   } & +66.67\%  & +834.61\% & -16.19\%  & +92.31\% \\ 
\textbf{Filmtrust   } & -1.01\%  & +10.64\% & -8.05\%  & +4.25\% \\ 
\textbf{Last.FM     } & -8.29\%  & +213.64\% & -15.39\%  & -22.22\% \\ 
\midrule
\textbf{Average     } & \textbf{+13.29\%}  & \textbf{+274.14\%} & \textbf{-27.42\%}  & \textbf{+53.41\%} \\ 
\textbf{Median      } & \textbf{-0.82\%}  & \textbf{+163.71\%} & \textbf{-17.51\%}  & \textbf{+34.89\%} \\ \bottomrule
\end{tabular}
\caption{Relational difference of Interact2Vec in NDCG@15 and elapsed training time compared to Item2Vec and RecVAE.}
\label{tab:itm2v_vs_int2v}
\end{table}

\begin{figure}[!htpb]
    \centering
    \includegraphics[width=0.85\linewidth]{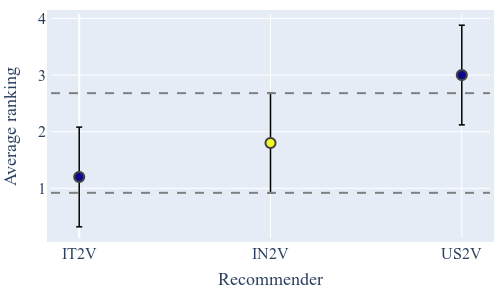}
    \caption{Graphical representation of the \textit{post-hoc} Bonferroni-Dunn test (embedding-based models only).}
    \label{fig:bonferroni_implicit_emb}
\end{figure}

To comprehensively compare Interact2Vec and the other neural embedding-based approaches, we performed a statistical test using a ranking constructed with the NDCG@$15$ results. Initially, a nonparametric Friedman~\cite{Demsar2006} test was conducted to assess whether significant differences exist among the performance of the methods. Results indicated that the models are statistically different, with 95\% confidence interval ($X^2_r = 16.80$). We then conducted a post-hoc Bonferroni-Dunn test~\cite{Demsar2006} to perform a pair-wise comparison between Interact2Vec, Item2Vec, and User2Vec (Figure~\ref{fig:bonferroni_implicit_emb}). It is possible to state that Interact2Vec is statistically superior to User2Vec, with a critical difference of $0.88$, and that there is no statistical evidence of the superiority of Interact2Vec to Item2Vec. The main advantages of the proposed model, when compared with Item2Vec, lie in its efficiency, being able to learn user and item embeddings simultaneously much faster than its counterpart (Figure~\ref{fig:itm2v_vs_int2v_elapsed_time}).

Although the performance of neural embedding-based methods is the main focus of comparison with Interact2Vec, we also conducted a similar test including the other recommender approaches to position the performance of Interact2Vec against other well-known strategies. Friedman's non-parametric test revealed a statistically significant difference between the models ($X^2_r = 33.39$), justifying the application of a Wilcoxon signed-rank test and the Bonferroni-Dunn test~\cite{Demsar2006}. 

\begin{figure}[!htpb]
    \centering
    \includegraphics[width=0.85\linewidth]{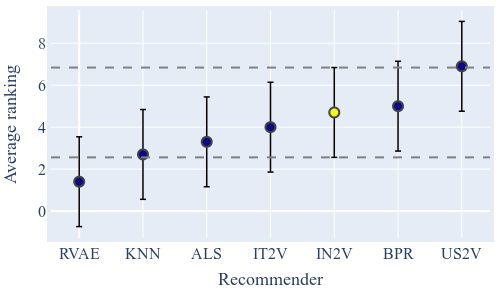}
    \caption{Graphical representation of the \textit{post-hoc} Bonferroni-Dunn test (all models).}
    \label{fig:bonferroni_implicit_all}
\end{figure}

According to the \textit{p}-values obtained by the two-tailed Wilcoxon signed-rank test (Table~\ref{tab:wilcoxon}), Interact2Vec is proven to be statistically different from KNN, User2Vec, and RecVAE with 95\% confidence interval. Considering the average ranking for the models, we can conclude that while Interact2Vec is inferior to KNN and RecVAE, the embedding-based model is statistically superior to User2Vec. This conclusion was partially maintained for the Bonferroni-Dunn test (Figure~\ref{fig:bonferroni_implicit_all}). Considering the critical difference determined by the test ($CD = 2.14$), we can confidently assert that Interact2Vec exhibits statistical superiority over User2Vec and is inferior to RecVAE. Although the Wilcoxon test indicates that KNN significantly outperforms Interact2Vec, their ranking difference did not reach the critical threshold in the Bonferroni-Dunn test when correcting for multiple comparisons -- a more conservative approach -- suggesting that while KNN tends to perform better, the improvement should be interpreted with caution. Nonetheless, while no statistically significant difference was found between Interact2Vec and the other models, further investigation may be necessary to understand the subtle advantages of its ranking compared to them.

\begin{table}[!htb]
\centering
\begin{tabular}{ccccccc}
    \toprule
    \textbf{Model} & \textbf{KNN} & \textbf{ALS} & \textbf{BPR} & \textbf{IT2V} & \textbf{US2V} & \textbf{RVAE} \\ \midrule
    \textit{p}-values & 0.0466 & 0.33204 & 0.80258 & 0.28462 & 0.00512 & 0.00512 \\ \bottomrule
\end{tabular}
\caption{\textit{p}-values for the Wilcoxon signed-rank test with 0.05 significance level.}
\label{tab:wilcoxon}
\end{table}

In conclusion, we cannot state that the performance obtained by Interact2Vec is statistically superior to other recommender models, such as KNN, ALS, and Item2Vec. However, when compared to other neural embedding models, the proposed approach is computationally much more efficient, presenting a pseudo-linear computational complexity. While User2Vec shares the same characteristic and is also capable of learning item and user embeddings simultaneously during training, its results were less satisfactory. In contrast, our proposed method, Interact2Vec, exhibits low computational complexity without significant compromises in the quality of the final recommendation. As a result, Interact2Vec stands out as a promising alternative for neural embedding models in recommender systems based on implicit feedback. The effectiveness of Interact2Vec makes it a viable choice in practical situations in which efficient methods are required, such as scenarios characterized by computational limitations.

\subsection{Similarity tables}\label{sec:similarity_table}

We conducted a subjective approach using similarity tables to evaluate the embeddings intrinsically, i.e., how well the vector representation retains knowledge of the items in its content. For this, we have analyzed models whose primary goal is learning user and item embeddings in the same vector space, thus focusing our comparisons on matrix factorization (ALS and BPR) and embedding-based models (IT2V, US2V, and IN2V).

Similarity tables are the easiest and most intuitive way to check if there is intrinsic meaning in the evaluated representation, being widely used in the literature~\cite{Barkan2016,Fu2017}. It consists of retrieving a subset of items, as well as their closest neighbors, and comparing if there is agreement in their content information. Thus, it is a strategy that relies heavily on human interpretation~\cite{Gladkova2016}.

To conduct an intrinsic evaluation of Interact2Vec and compare it with the other vector representation methods, four popularly known items within the Last.FM and MovieLens data were arbitrarily selected. These datasets were chosen because they contain items from popular domains (music and films, respectively). For each of the four selected items, the three items closest to them in the embedding vector space were retrieved for each recommendation method. The results for the Last.FM dataset are found in Table~\ref{tab:lastfm_sims}, and for the MovieLens database in Table~\ref{tab:movielens_sims}. For each selected artist or film, their musical or film genres were indicated, enabling a greater background of the items.

\begin{table*}[!ht]
    \centering
    \begin{tabularx}{\textwidth}{XXXXXX}
    \toprule
     \multirow{2}{*}{\textbf{Seed}} & \multicolumn{5}{c}{\textbf{Representation model}} \\ \cline{2-6}
     & \textbf{ALS} & \textbf{BPR} & \textbf{IT2V} & \textbf{US2V} & \textbf{IN2V}\\ \midrule
     
      & \scriptsize{medusa' Scream} \newline \textit{\scriptsize{emo}} & \scriptsize{30 Seconds to Mars} \newline \textit{\scriptsize{rock, alt rock}} & \scriptsize{Breaking} \newline \scriptsize{Benjamin} \newline \textit{\scriptsize{rock, alt rock}} & \scriptsize{30 Seconds to Mars} \newline \textit{\scriptsize{rock, alt rock}} & \scriptsize{Marilyn} \newline \scriptsize{Manson} \newline \textit{\scriptsize{metal, indust.}} \\ \cline{2-6}
      \scriptsize{Linkin Park} \newline \textit{\scriptsize{alt rock, rock}} & \scriptsize{Grey Daze} \newline \textit{\scriptsize{alt rock, rock}} & \scriptsize{The Rasmus} \newline \textit{\scriptsize{rock, alt}} & \scriptsize{30 Seconds to Mars} \newline \textit{\scriptsize{rock, alt rock}} & \scriptsize{Flyleaf} \newline \textit{\scriptsize{alt rock, rock}} & \scriptsize{Nickelback} \newline \textit{\scriptsize{rock,}} \newline \textit{\scriptsize{hard rock}} \\ \cline{2-6}
      & \scriptsize{Dead by} \newline \scriptsize{Sunrise} \newline \textit{\scriptsize{alt rock, rock}} & \scriptsize{Breaking} \newline \scriptsize{Benjamin} \newline \textit{\scriptsize{rock, alt rock}} & \scriptsize{Evanescence} \newline \textit{\scriptsize{rock, female}} & \scriptsize{VersaEmerge} \newline \textit{\scriptsize{rock, female}} & \scriptsize{30 Seconds to Mars} \newline \textit{\scriptsize{rock, alt rock}} \\ \hline
    
     & \scriptsize{Juanes} \newline \textit{\scriptsize{latin, pop}} & \scriptsize{Beyoncé} \newline \textit{\scriptsize{rnb, pop}} & \scriptsize{Rihanna} \newline \textit{\scriptsize{pop, rnb}} & \scriptsize{Katy Perry} \newline \textit{\scriptsize{pop, female}} & \scriptsize{Jennifer} \newline \scriptsize{Lopez} \newline \textit{\scriptsize{pop, dance}} \\ \cline{2-6}
     \scriptsize{Shakira} \newline \textit{\scriptsize{pop, female}} & \scriptsize{Fanny Lu} \newline \textit{\scriptsize{latin pop}} & \scriptsize{Marilyn} \newline \scriptsize{Monroe} \newline \textit{\scriptsize{jazz, female}} & \scriptsize{Beyoncé} \newline \textit{\scriptsize{rnb, pop}} & \scriptsize{Mariah} \newline \scriptsize{Carey} \newline \textit{\scriptsize{rnb, pop}} & \scriptsize{Miley Cyrus} \newline \textit{\scriptsize{pop, female}} \\ \cline{2-6}
     & \scriptsize{Thalía} \newline \textit{\scriptsize{female, latin}} & \scriptsize{Rihanna} \newline \textit{\scriptsize{pop, rnb}} & \scriptsize{Britney} \newline {Spears} \newline \textit{\scriptsize{pop, dance}} & \scriptsize{Beyoncé} \newline \textit{\scriptsize{rnb, pop}} & \scriptsize{Mariah} \newline \scriptsize{Carey} \newline \textit{\scriptsize{rnb, pop}} \\ \hline
    
      & \scriptsize{Ozzy} \newline \scriptsize{Osbourne} \newline \textit{\scriptsize{heavy metal}} & \scriptsize{Velvet} \newline \scriptsize {Revolver} \newline \textit{\scriptsize{hard rock, alt}} & \scriptsize{Van Halen} \newline \textit{\scriptsize{hard rock, 80s}} & \scriptsize{Heaven \& Hell} \newline \textit{\scriptsize{heavy metal}} & \scriptsize{Aerosmith} \newline \textit{\scriptsize{hard rock}} \newline \textit{\scriptsize{classic rock}} \\ \cline{2-6}
      \scriptsize{AC/DC} \newline \textit{\scriptsize{classic rock}} \newline \textit{\scriptsize{hard rock}} & \scriptsize{Led Zeppelin} \newline \textit{\scriptsize{classic rock}} \newline \textit{\scriptsize{hard rock}} & \scriptsize{Guns N Roses} \newline \textit{\scriptsize{hard rock,}} \newline \textit{\scriptsize{rock}} & \scriptsize{Metallica} \newline \textit{\scriptsize{metal}} \newline \textit{\scriptsize{thrash metal}} & \scriptsize{Metal Church} \newline \textit{\scriptsize{thrash metal}} & \scriptsize{Iron Maiden} \newline \textit{\scriptsize{heavy metal}} \newline \textit{\scriptsize{metal}} \\ \cline{2-6}
      & \scriptsize{Guns N Roses} \newline \textit{\scriptsize{hard rock,}} \newline \textit{\scriptsize{rock}} & \scriptsize{Tenacious D} \newline \textit{\scriptsize{rock,}} \newline \textit{\scriptsize{hard rock}} & \scriptsize{Guns N Roses} \newline \textit{\scriptsize{hard rock,}} \newline \textit{\scriptsize{rock}} & \scriptsize{Iommi} \newline \textit{\scriptsize{heavy metal}} \newline \textit{\scriptsize{metal}} & \scriptsize{Motörhead} \newline \textit{\scriptsize{heavy metal}} \newline \textit{\scriptsize{hard rock}} \\ \hline
    

     & \scriptsize{Ricky Nelson} \newline \textit{\scriptsize{rock,}} \newline \textit{\scriptsize{classic rock}} & \scriptsize{Beach Boys} \newline \textit{\scriptsize{60s}} \newline \textit{\scriptsize{classic rock}} & \scriptsize{David Bowie} \newline \textit{\scriptsize{rock}} \newline \textit{\scriptsize{classic rock}} & \scriptsize{The Kinks} \newline \textit{\scriptsize{60s}} \newline \textit{\scriptsize{classic rock}} & \scriptsize{Radiohead} \newline \textit{\scriptsize{alt, rock}} \\ \cline{2-6}
     \scriptsize{The Beatles} \newline \textit{\scriptsize{rock}} \newline \textit{\scriptsize{classic rock}} & \scriptsize{Souad Massi} \newline \textit{\scriptsize{female, arabic}} & \scriptsize{John Lennon} \newline \textit{\scriptsize{classic rock}} \newline \textit{\scriptsize{rock}} & \scriptsize{Radiohead} \newline \textit{\scriptsize{alt, rock}} & \scriptsize{The Rolling} \newline \scriptsize{Stones} \newline \textit{\scriptsize{classic rock}} & \scriptsize{Pink Floyd} \newline \textit{\scriptsize{prog rock}} \newline \textit{\scriptsize{classic rock}} \\ \cline{2-6}
     & \scriptsize{Andrés} \newline \scriptsize{Segovia} \newline \textit{\scriptsize{baroque}} & \scriptsize{Ringo Starr} \newline \textit{\scriptsize{classic rock}} \newline \textit{\scriptsize{rock}} & \scriptsize{Led Zeppelin} \newline \textit{\scriptsize{classic rock}} \newline \textit{\scriptsize{hard rock}} & \scriptsize{The Velvet} \newline \scriptsize{Underground} \newline \textit{\scriptsize{psychedelic}} & \scriptsize{Led Zeppelin} \newline \textit{\scriptsize{classic rock}} \newline \textit{\scriptsize{hard rock}} \\ \bottomrule
    \end{tabularx}
    \caption{Similarity table of four popular artists from the Last.FM dataset}
    \label{tab:lastfm_sims}
    \end{table*}

Although the analysis of the tables is highly biased by human opinions, we can note that, for the Last.FM database (Table~\ref{tab:lastfm_sims}), all models found items with genres similar to the targets, with the only exception of ALS to \textit{The Beatles}. However, it is not easy to assess whether there is any model that stands out from the others since, even though there is a considerable difference between the related items, they all have some relationship with the target item.

Except for ALS, User2Vec was the model that most recommended little-known artists, such as \textit{Flyleaf} or \textit{VersaEmerge} for \textit{Pink Floyd}, and \textit{Heaven \& Hell} or \textit {Metal Church} for \textit{AC/DC}. Still, there is an agreement regarding the musical genre between the recommended artists and the target item, making it difficult to compare the quality of their representation. ALS was another method that recommended less popular artists, creating a curious situation for \textit{Shakira}: just like the Colombian artist, all other methods recommended world-famous pop singers, such as \textit{Beyoncé} and \textit{Mariah Carey}, possibly indicating better representation. However, all the artists recommended by ALS are pop singers of Latin nationality, having an additional feature related to the target that is not shared by the other recommendations. Therefore, it is a complex task to evaluate the quality of representations, which is extremely linked to the evaluator's knowledge of the domain.

\begin{table*}[!htpb]
\centering
\begin{tabularx}{\textwidth}{XXXXXX}
\toprule
 \multirow{2}{*}{\textbf{Seed}} & \multicolumn{5}{c}{\textbf{Representation model}} \\ \cline{2-6}
 & \textbf{ALS} & \textbf{BPR} & \textbf{IT2V} & \textbf{US2V} & \textbf{IN2V}\\ \midrule

 & \scriptsize{Average} \newline \scriptsize {Italian} \newline \textit{\scriptsize{Comedy}} & \scriptsize{Muppet Treasure Island} \newline \textit{\scriptsize{Children}} & \scriptsize{Braveheart} \newline \textit{\scriptsize{Drama, War}} & \scriptsize{Lion King} \newline \textit{\scriptsize{Children,}} \newline \textit{\scriptsize{Musical}} & \scriptsize{Star Wars IV} \newline \textit{\scriptsize{Adventure,}} \newline \textit{\scriptsize{Sci-Fi}} \\ \cline{2-6}
\scriptsize{Toy Story} \newline \textit{\scriptsize{Children,}}  \newline \textit{\scriptsize{Comedy}} & \scriptsize{The Pride} \newline {and Passion} \newline \textit{\scriptsize{War}} & \scriptsize{Babe} \newline \textit{\scriptsize{Children, Drama}} & \scriptsize{12 Monkeys} \newline \textit{\scriptsize{Sci-Fi,}} \newline \textit{\scriptsize{Thriller}} & \scriptsize{Toy Story 2} \newline \textit{\scriptsize{Children,}} \newline \textit{\scriptsize{Comedy}} & \scriptsize{LOTR II} \newline \textit{\scriptsize{Adventure, Fantasy}} \\ \cline{2-6}
 & \scriptsize{Barbie} \newline \textit{\scriptsize{Animation, Children}} & \scriptsize{Jack-O} \newline \textit{\scriptsize{Horror}} & \scriptsize{The Usual} \newline \scriptsize{Suspects} \newline \textit{\scriptsize{Crime}} & \scriptsize{Men in Black} \newline \textit{\scriptsize{Action, Sci-Fi}} & \scriptsize{Sixth Sense} \newline \textit{\scriptsize{Drama,}} \newline \textit{\scriptsize{Horror}} \\ \hline

 & \scriptsize{Dirty} \newline \scriptsize{Dancing} \newline \textit{\scriptsize{Musical}} & \scriptsize{The Sound of Music} \newline \textit{\scriptsize{Musical}} & \scriptsize{Batman Rtns.} \newline \textit{\scriptsize{Action, Crime}} & \scriptsize{Top Gun} \newline \textit{\scriptsize{Action,}} \newline \textit{\scriptsize{Romance}} & \scriptsize{First Knight} \newline \textit{\scriptsize{Action}} \newline \textit{\scriptsize{Romance}} \\ \cline{2-6}
\scriptsize{Grease} \newline \textit{\scriptsize{Musical,}}  \newline \textit{\scriptsize{Romance}} & \scriptsize{Big} \newline \textit{\scriptsize{Comedy,}} \newline \textit{\scriptsize{Romance}} & \scriptsize{The Little} \newline \scriptsize{Princess} \newline \textit{\scriptsize{Children}} & \scriptsize{Scream} \newline \textit{\scriptsize{Horror,}} \newline \textit{\scriptsize{Mystery}} & \scriptsize{Gremlins} \newline \textit{\scriptsize{Comedy}} \newline \textit{\scriptsize{Horror}} & \scriptsize{Deep Impact} \newline \textit{\scriptsize{Drama, Sci-Fi}} \\ \cline{2-6}
 & \scriptsize{The Little} \newline \scriptsize{Mermaid} \newline \textit{\scriptsize{Musical}} & \scriptsize{Oliver!} \newline \textit{\scriptsize{Drama,}} \newline \textit{\scriptsize{Musical}} & \scriptsize{Air Force 1} \newline \textit{\scriptsize{Action,}} \newline \textit{\scriptsize{Thriller}} & \scriptsize{Mary} \newline \scriptsize{Poppins} \newline \textit{\scriptsize{Musical}} & \scriptsize{Sneakers} \newline \textit{\scriptsize{Action, Crime}} \\ \hline


 & \scriptsize{Groundhog Day} \newline \textit{\scriptsize{Comedy}} & \scriptsize{The Truman} \newline \scriptsize{Show} \newline \textit{\scriptsize{Comedy}} & \scriptsize{Good Will} \newline \scriptsize{Hunting} \newline \textit{\scriptsize{Drama}} & \scriptsize{Jurassic Park} \newline \textit{\scriptsize{Action, Sci-Fi}} & \scriptsize{The Usual} \newline \scriptsize{Suspects} \newline \textit{\scriptsize{Crime}} \\ \cline{2-6}
\scriptsize{Titanic} \newline \textit{\scriptsize{Drama,}}  \newline \textit{\scriptsize{Romance}} & \scriptsize{The Truman} \newline \scriptsize{Show} \newline \textit{\scriptsize{Comedy}} & \scriptsize{Catch Me If You Can} \newline \textit{\scriptsize{Drama}} & \scriptsize{Men in Black} \newline \textit{\scriptsize{Action, Sci-Fi}} & \scriptsize{The Truman} \newline \scriptsize{Show} \newline \textit{\scriptsize{Comedy}} & \scriptsize{Blade Runner} \newline \textit{\scriptsize{Action, Sci-Fi}} \\ \cline{2-6}
 & \scriptsize{Christmas Do-Over} \newline \textit{\scriptsize{Comedy}} & \scriptsize{Best Friends} \newline \scriptsize{Wedding} \newline \textit{\scriptsize{Comedy}} & \scriptsize{Saving Private Ryan} \newline \textit{\scriptsize{Drama, War}} & \scriptsize{Men in Black} \newline \textit{\scriptsize{Action, Sci-Fi}} & \scriptsize{Indiana} \newline \scriptsize{Jones I} \newline \textit{\scriptsize{Adventure}} \\ \hline

& \scriptsize{A View to a Kill} \newline \textit{\scriptsize{Action}} & \scriptsize{Nightmare on Elm Street 4} \newline \textit{\scriptsize{Horror}} & \scriptsize{Gremlins 2} \newline \textit{\scriptsize{Comedy}} \newline \textit{\scriptsize{Horror}} & \scriptsize{Friday the} \newline \scriptsize{13th Part 2} \newline \textit{\scriptsize{Horror}} & \scriptsize{Jaal: The} \newline \scriptsize{Trap} \newline \textit{\scriptsize{-}} \\ \cline{2-6}
\scriptsize{Friday the} \newline \scriptsize{13th} \newline  \textit{\scriptsize{Horror}} & \scriptsize{Child's Play} \newline \textit{\scriptsize{Horror,}} \newline \textit{\scriptsize{Thriller}} & \scriptsize{Friday the} \newline \scriptsize{13th Part 3} \newline \textit{\scriptsize{Horror}} & \scriptsize{Texas Chainsaw Mass.} \newline \textit{\scriptsize{Horror}} & \scriptsize{Halloween II} \newline \textit{\scriptsize{Horror}} & \scriptsize{Calcutta Mail} \newline \textit{\scriptsize{Thriller}} \\ \cline{2-6}
 & \scriptsize{Pet Sematary} \newline \textit{\scriptsize{Horror}} & \scriptsize{Children of} \newline \scriptsize{the Corn} \newline \textit{\scriptsize{Horror}} & \scriptsize{Halloween} \newline \textit{\scriptsize{Horror}} & \scriptsize{Child's Play} \newline \textit{\scriptsize{Horror,}} \newline \textit{\scriptsize{Thriller}} & \scriptsize{Saaya} \newline \textit{\scriptsize{Thriller,}} \newline \textit{\scriptsize{Romance}} \\ \bottomrule
\end{tabularx}
\caption{Similarity table of four popular movies from the MovieLens dataset}
\label{tab:movielens_sims}
\end{table*}

For the movies dataset (Table~\ref{tab:movielens_sims}), the results between the methods showed a greater discrepancy. The items recommended by Interact2Vec exhibited significant divergence from the target item, such as the choice of action films for \textit{Grease} (a romantic musical) or productions of Indian origin for \textit{Friday the 13th} (an American horror film). Given the suboptimal performance of Interact2Vec in the top-$N$ ranking task for the MovieLens dataset, these observed deviations are anticipated and underscore the evidence that the proposed model faced challenges in learning a representative embedding for the cinematic dataset.

Upon closer examination of the other methods, it is difficult to select one that has presented superior results. The observed similarities between films are often not exclusively rooted in their genres, as demonstrated by instances where movies are deemed similar due to their status as cinematic classics, such as \textit{Titanic} and \textit{Men in Black}, or \textit{Saving Private Ryan}. Thus, there are favorable arguments for almost all choices made by the methods. Another interesting point of analysis is how certain methods presented extremely accurate results for some films and questionable results for others, such as BPR: the method was the only one that recommended three musicals for \textit{Grease}, but it also recommended one horror film (\textit{Jack-O}) to the children's animation \textit{Toy Story}.

The subjective evaluation adopted in this work may lead to inconsistencies in the observed results, as such assessments are inherently influenced by individual perception and contextual factors. This makes it challenging to select the best-performing model definitively. It is worth noting that this issue was explored in greater depth, focusing on traditional embedding-based recommenders~\cite{Pires2024}. Their findings highlight how subjective interpretations can shape the perceived quality of learned representations. While a deeper investigation lies beyond the scope of the present work, these observations reinforce the need for continued research on extrinsic evaluation strategies for recommender systems.

\section{Conclusions and future work}

In this paper, we introduced Interact2Vec, a novel neural-based approach designed to simultaneously learn user and item embeddings for recommender systems. Interact2Vec offers key advantages, including high computational efficiency and independence from content data, making it adaptable to a wide range of application scenarios.

We evaluated Interact2Vec in a top-$N$ recommendation task across multiple datasets, benchmarking it against other neural embedding-based approaches such as Item2Vec and User2Vec. Interact2Vec consistently outperformed User2Vec, achieving superior F1 scores and NDCG in every experiment. While its performance was comparable to Item2Vec, Interact2Vec demonstrated a significant advantage in computational performance, with an average training time reduction of 274\%.

Although the performance of neural embedding-based methods is the main focus of comparison, to further contextualize the efficacy of Interact2Vec, we compared it against well-established approaches such as KNN, ALS, and BPR, as well as a state-of-the-art deep learning model, RecVAE. Our approach achieved competitive results with traditional algorithms but underperformed compared to RecVAE. However, Interact2Vec offers substantial benefits over these models, particularly in generating low-dimensional representations in a computationally efficient manner.

Additionally, we conducted an intrinsic evaluation to explore whether the model could learn embeddings that encapsulate content-related information without explicitly consuming content data. This analysis, involving similarity tables and an automatic feature prediction task, revealed that Interact2Vec struggled in learning content information, particularly for infrequently consumed items. In contrast, Item2Vec and User2Vec were better suited for applications requiring intrinsic knowledge.

Since Interact2Vec does not consume explicit feedback, i.e., user ratings, it is suitable for top-$N$ recommendations. For future work, we propose investigating methods to extend the model's capabilities to handle explicit feedback, enabling it to perform rating prediction tasks. Two promising directions include (\textit{i}) modifying the objective function to minimize prediction error based on ratings and (\textit{ii}) introducing new similarity weighting techniques that factor in the user's ratings.

In future research, we also plan to explore how different recommender algorithms influence recommendation quality. Specifically, we aim to identify correlations between algorithm type and the characteristics of the consumed data. In addition, based on existing literature~\cite{Vasile2016,Fu2017}, incorporating content data could improve embedding quality, especially in tasks demanding intrinsic knowledge, though this would introduce complexity to the model. We also foresee potential improvements by integrating temporal information into the embeddings, as other neural embedding models have shown enhanced performance using this strategy~\cite{Koren2009b,Pires2021}.

Finally, in future work, we plan to include comparisons between Interact2Vec and emerging state-of-the-art recommenders that, while not necessarily focused on embedding learning, offer high computational efficiency, e.g., LightGCN~\cite{He2020} and SimRec~\cite{Brody2024}, to validate the practical competitiveness of our approach further. We also recommend conducting a more comprehensive study to evaluate the intrinsic quality of embeddings generated by Interact2Vec compared to other embedding-based methods. While this study focused on extrinsic evaluation through the lens of recommendation performance, a deeper understanding of Interact2Vec's capabilities could be achieved by testing the embeddings in additional tasks such as auto-tagging and item clustering.


\appendix

\section{Sensitivity evaluation of Interact2Vec parameters}
\label{apx:in2v_parameters}

Like other embedding-based models, Interact2Vec has many hyperparameters that can be adjusted before learning. This means that parameter optimization techniques that perform exhaustive tests, such as grid search, may be unfeasible, generating several combinations of values.

\begin{table}[htpb]
    \centering
    \begin{tabular}{@{}lcc@{}}
        \toprule
        \textbf{Hyperparameter} & \textbf{Notation} & \textbf{Default value} \\ \midrule
        Learning rate & $\alpha$ & 0.25 \\ 
        Embedding size & $M$ & 100 \\ 
        Number of epochs & $C$ & 5 \\ 
        Subsampling rate of frequent items & $\rho$ & $10^{-3}$ \\ 
        Number of negative samples & $|G|$ & 5 \\ 
        Exponent of the negative sampling & $\gamma$ & 0.75 \\ 
        Regularization factor & $\lambda$ & 0 \\
        \bottomrule
    \end{tabular}
    \caption{Commonly used values when learning neural embeddings.}
    \label{tab:parametrosBaseOtimizacao}
\end{table}

To alleviate this problem, we conducted an impact study for Interact2Vec parameters over datasets CaioDVD and Last.FM, since they are from different domains and do not have a high volume of data, thus enabling the execution of several models quickly. The following experiment was conducted: a default value was defined for each parameter, normally used by other models~\cite{Barkan2016} or libraries for generating embeddings~\cite{Rehurek2010} (Table~ \ref{tab:parametrosBaseOtimizacao}). Then, the datasets were separated into training, validation, and testing, following an 8:1:1 rate, and several variations of the proposed model were adjusted over training and evaluated on a validation basis. The content of each parameter analyzed was varied within an extensive list of values, with all other parameters fixed at standard values. To evaluate each model, the NDCG@15 was used.

The following sections analyze how each parameter impact on the final recommendation.

\subsection{Learning rate}

\begin{figure}[htb]
    \centering
    \includegraphics[width=0.8\linewidth]{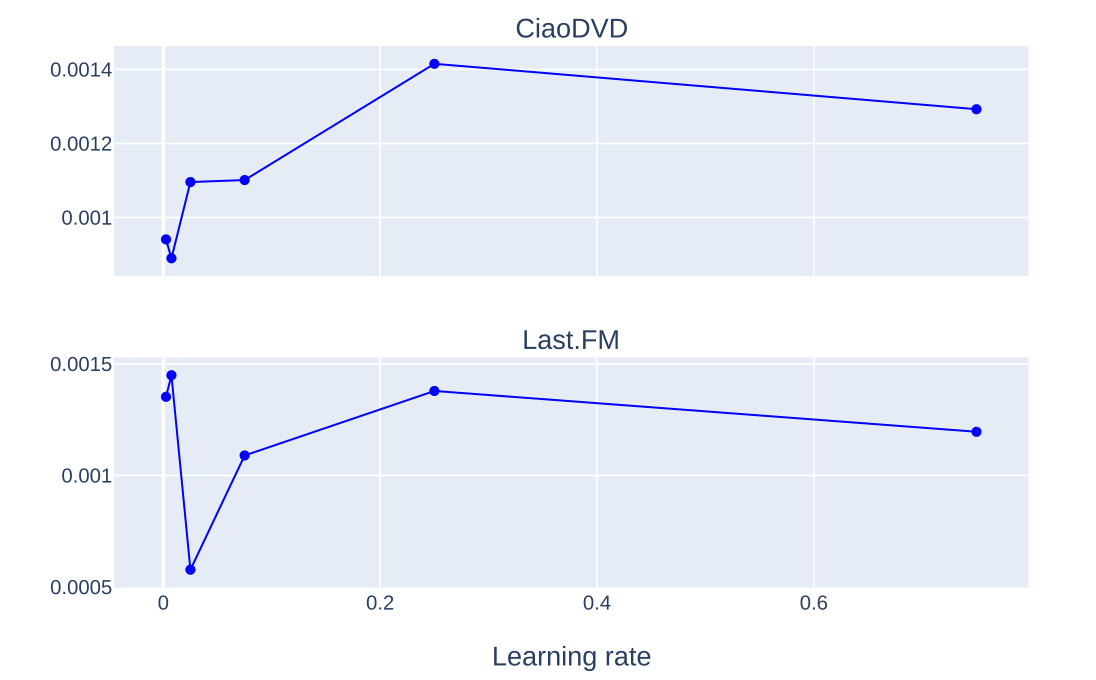}
    \caption{NDCG@15 for different learning rate values ($\alpha$).}
    \label{apx-fig:param_lr}
\end{figure}

We optimized the weight matrices of Interact2Vec using the Adam optimizer algorithm. We tested different values for the learning rate $\alpha$, ranging in $\{0.0025,\;0.0075,\;0.025,\;0.075,\;0.25,\;0.75\}$. Results can be seen in Figure~\ref{apx-fig:param_lr}, showing how worse NDCG values were achieved when $\alpha$ is smaller. In both datasets, the best results were achieved when $\alpha=0.25$, which was then selected as the default value.

\subsection{Embedding size}

\begin{figure}[htb]
    \centering
    \includegraphics[width=0.8\linewidth]{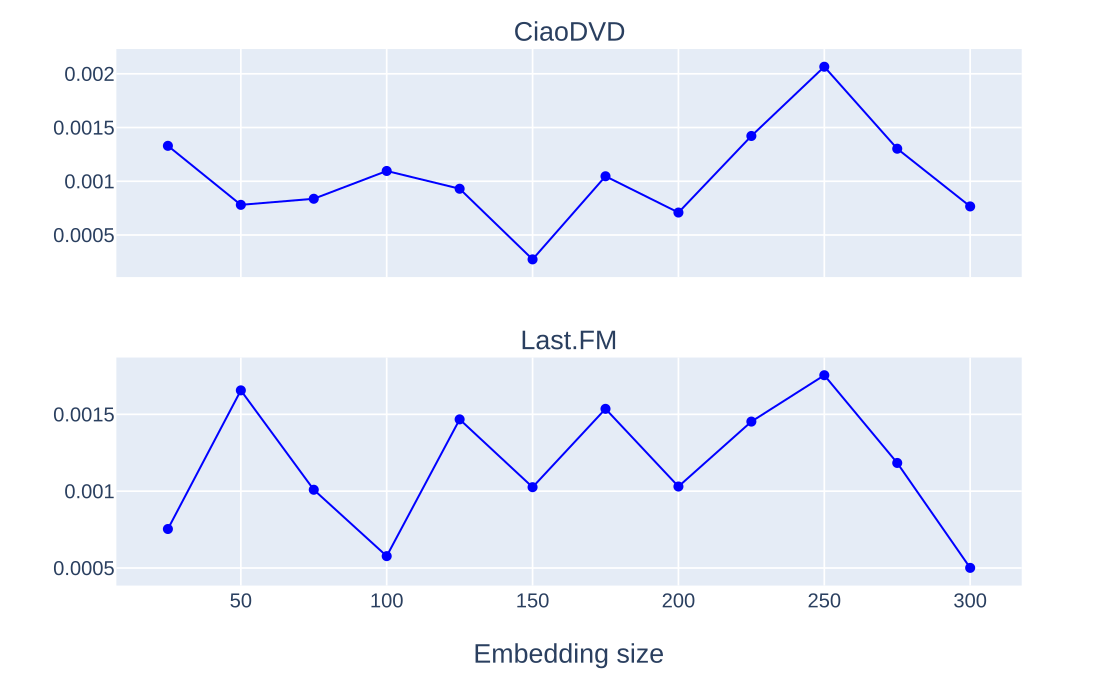}
    \caption{NDCG@15 for different embedding size values ($M$).}
    \label{apx-fig:param_emb_size}
\end{figure}

The embedding final size $M$ is a hyperparameter that directly impacts the overall efficiency of the recommender. In many cases, the choice of $M$ can be guided by the application domain's memory and processing power constraints. Even so, it is important to analyze if there are specific values of $M$ that tend to perform better.

We tested multiple embedding sizes, ranging in the interval $[25,300]$ with steps of 25. Results can be seen in Figure~\ref{apx-fig:param_emb_size}. The large variation in the NDCG, especially for the Last.FM dataset, highlights how different values for $M$ can change the quality of the representation, but with no well-defined pattern of better values. Selecting an appropriate value for the size of the embeddings can improve the result by 60\% or more, as seen for the CiaoDVD dataset, so it is a hyperparameter that is worth tuning.

\subsection{Number of epochs}

\begin{figure}[htb]
    \centering
    \includegraphics[width=0.8\linewidth]{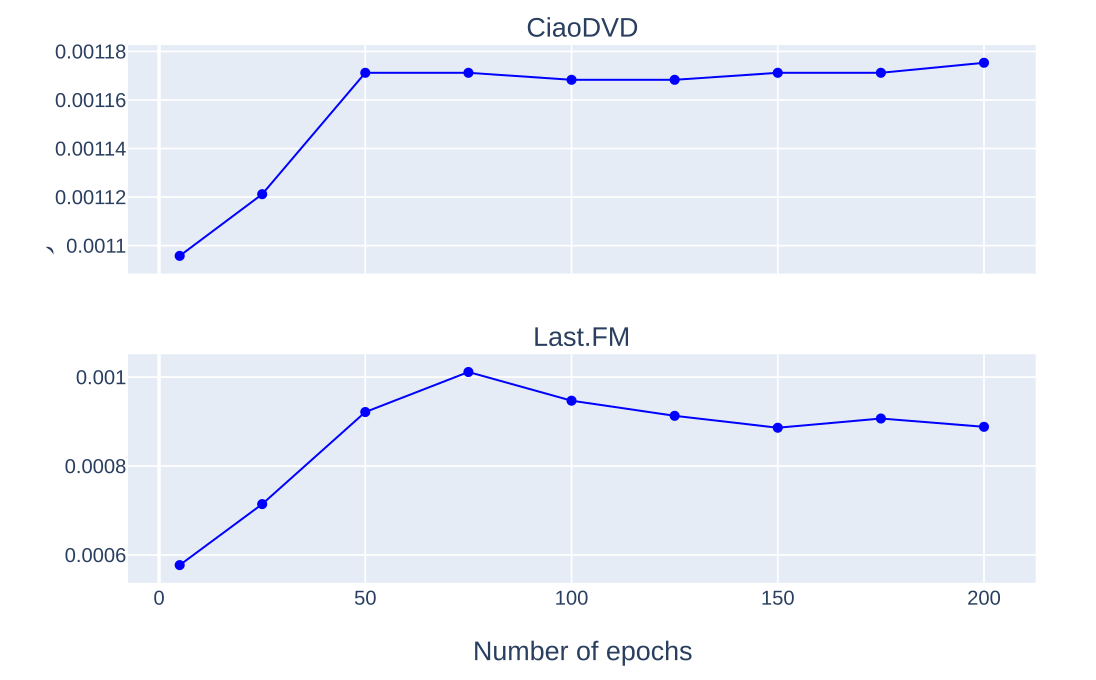}
    \caption{NDCG@15 for different number of epochs ($C$).}
    \label{apx-fig:param_n_epochs}
\end{figure}

For the number of epochs $C$ when training the neural networks, we tested values in the interval of $[5,200]$ with steps of 25. Results are presented in Figure~\ref{apx-fig:param_n_epochs}.

As observed, experiments with fewer epochs presented a lower NDCG@15 than experiments where $C >= 50$. On the other hand, the metric did not undergo major changes for values greater than $50$, showing only a slight drop for the Last.FM dataset. With this in mind, the values of $C$ were not fine-tuned during the experiments, with 50 used as default. 

It is important to highlight that the value normally used by neural embedding-based model libraries ($C = 5$) tends to present a lower result. Therefore, if the parameter is not adjusted, it is interesting to set it between $50$ and $100$ epochs as default.

\subsection{Subsampling rate of frequent items}

\begin{figure}[htb]
    \centering
    \includegraphics[width=0.8\linewidth]{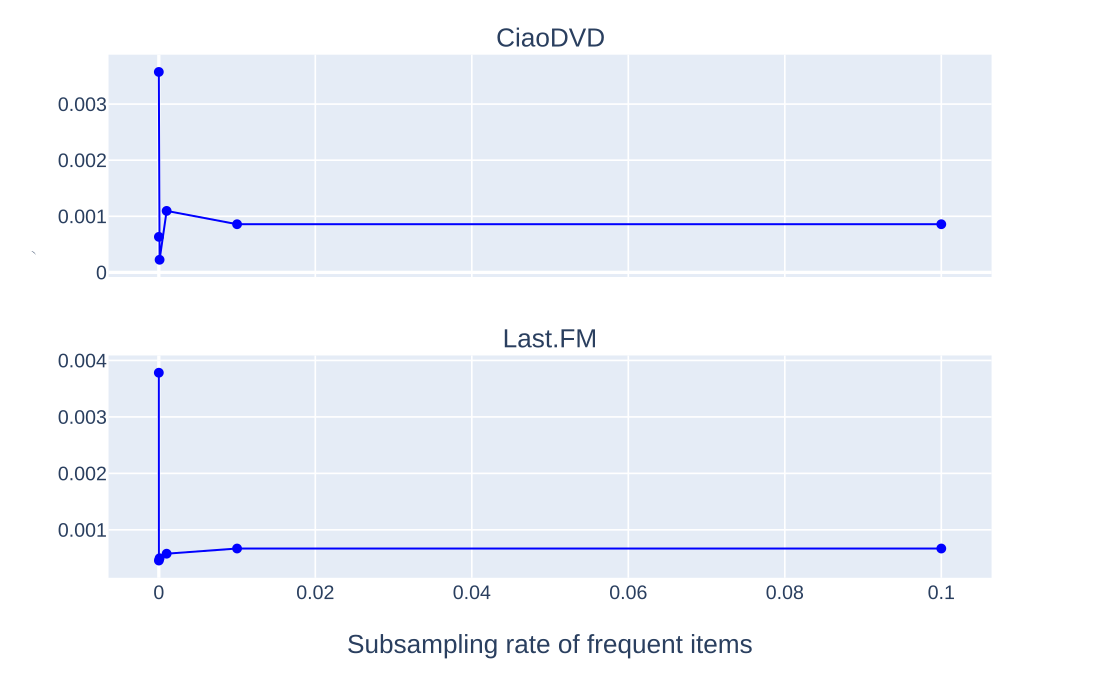}
    \caption{NDCG@15 for different frequent item subsampling rate values ($\rho$).}
    \label{apx-fig:param_sub_p}
\end{figure}

For the hyperparameter $\rho$, which controls the subsampling rate when discarding interactions of frequent items, we tested values ranging in $[10^{-6}, 10^{-1}]$ with step $\times 10$. Results are presented in Figure~\ref{apx-fig:param_sub_p} and show how using low values for $\rho$ improves the results significantly in both datasets. We then adopted $10^{-6}$ as a default value for the parameter when conducting the experiments.

\subsection{Number of negative samples}
\begin{figure}[!htb]
    \centering
    \includegraphics[width=0.8\linewidth]{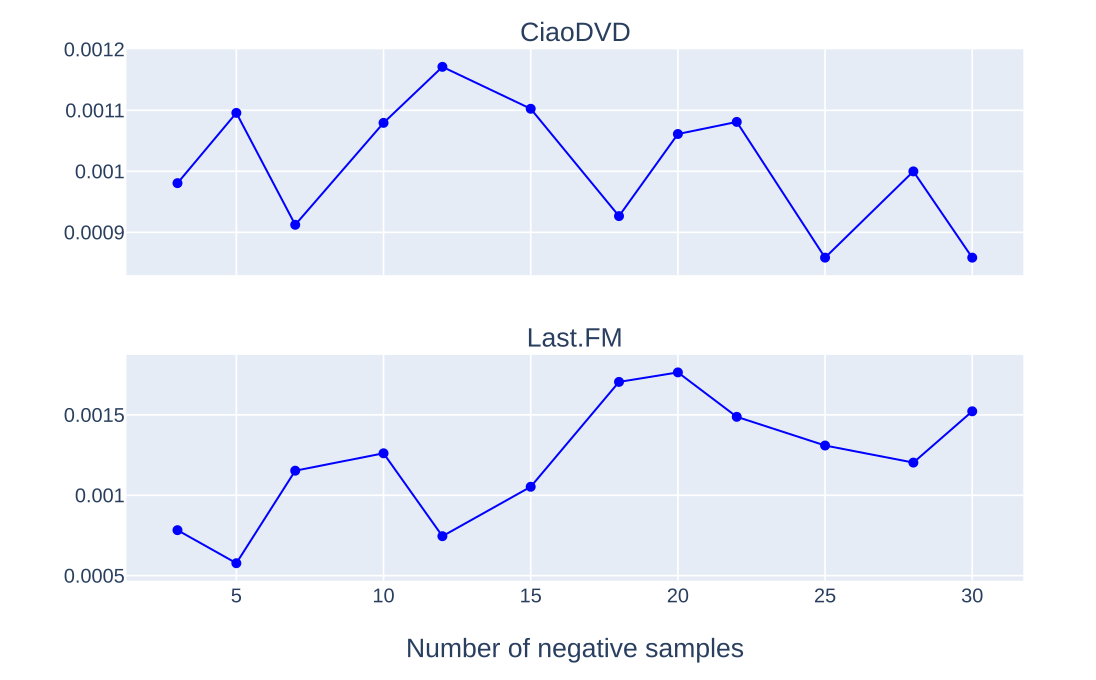}
    \caption{NDCG@15 for different numbers of negative samples ($|G|$).}
    \label{apx-fig:param_neg_samp}
\end{figure}

For the number $|G|$ of negative samples drawn at each positive sample iteration, we tested values ranging in the interval $[3,30]$ with steps 2 and 3 alternately. Results can be seen in Figure~\ref{apx-fig:param_neg_samp}.

There was a significant variation in the NDCG values, with no consensus on a more appropriate default value. For the Last.FM dataset, for example, different values for $|G|$ can change the result by more than 50\%, thus becoming an interesting parameter to be adjusted.

\subsection{The exponent of the negative sampling}

\begin{figure}[!htb]
    \centering
    \includegraphics[width=0.8\linewidth]{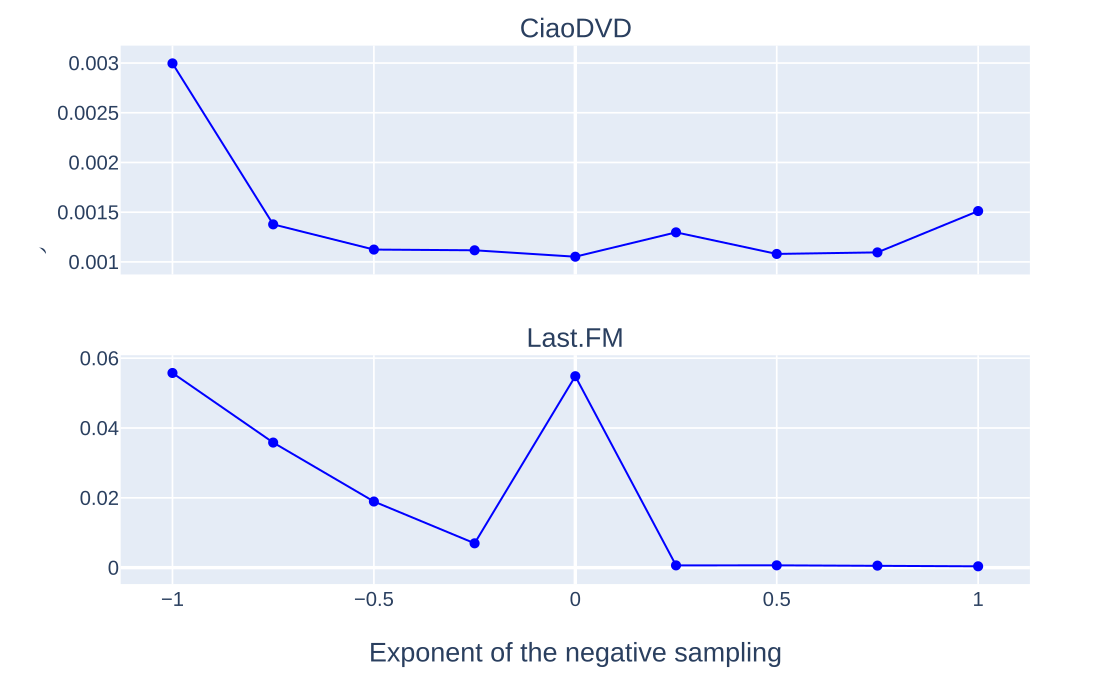}
    \caption{NDCG@15 for different values of the exponent for the negative sampling ($\gamma$).}
    \label{apx-fig:param_neg_distr}
\end{figure}

The exponent of the negative sampling $\gamma$ controls the probability of selecting negative samples. Here, we tested values inside the interval $[-1.0,\;1.0]$ with steps of $0.25$, with results presented in Figure~\ref{apx-fig:param_neg_distr}.

\begin{figure}[!htb]
    \centering
    \includegraphics[width=0.8\linewidth]{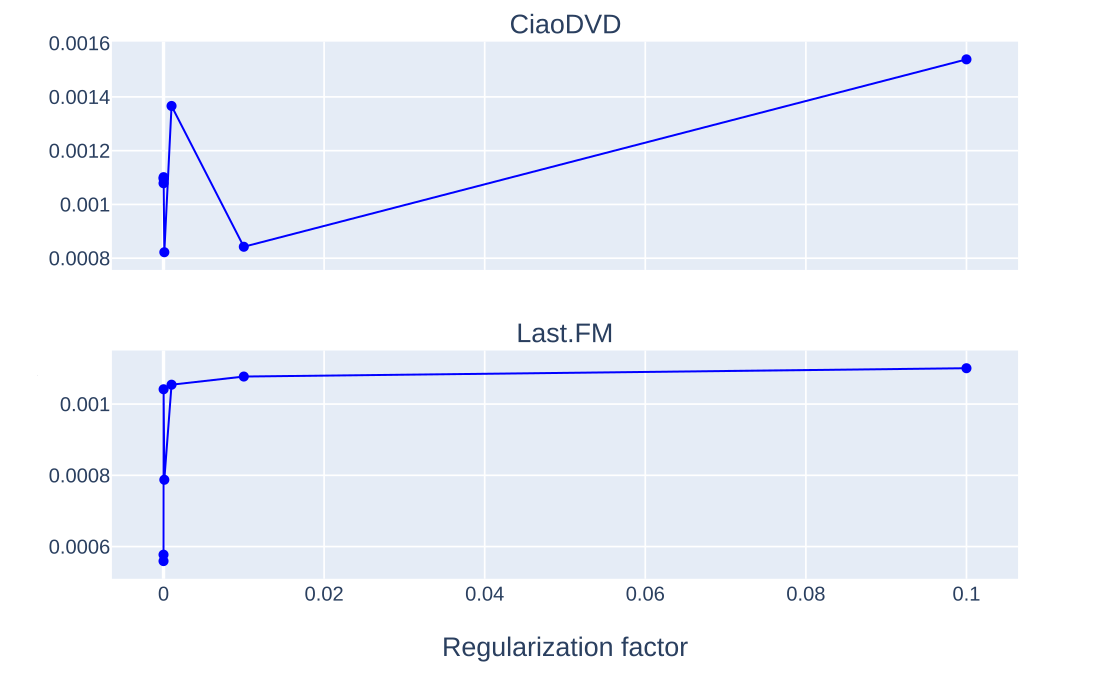}
    \caption{NDCG@15 for different values of the regularization factor ($\lambda$).}
    \label{apx-fig:param_reg}
\end{figure}

The exponent $\gamma$ was the parameter that showed the greatest variation among all the parameters evaluated in the experiment. For Last.FM, different values for the parameter can result in an improvement of 5849.19\%. Although there is evidence that negative values behave better, as seen in both datasets, a variation of this intensity makes $\gamma$ a mandatory parameter for fine-tuning.

\subsection{Regularization factor}

Finally, for the regularization factor $\lambda$ responsible for alleviating the problem of overfitting, we tested values ranging in $[10^{-6}, 10^{-1}]$, with steps of $(\times 10)$. We also tested $\lambda = 0$, i.e., without using regularization. Results can be seen in Figure~\ref{apx-fig:param_reg}.

For the two selected datasets, the best result was found when $\lambda = 10^{-1}$. This indicates an interesting behavior and a value that can be adopted as default in scenarios where there is no possibility of adjusting the parameter.

 \bibliographystyle{elsarticle-num} 
 \bibliography{references}

\begin{thebibliography}{10}
\expandafter\ifx\csname url\endcsname\relax
  \def\url#1{\texttt{#1}}\fi
\expandafter\ifx\csname urlprefix\endcsname\relax\def\urlprefix{URL }\fi
\expandafter\ifx\csname href\endcsname\relax
  \def\href#1#2{#2} \def\path#1{#1}\fi

\bibitem{Bobadilla2013}
J.~Bobadilla, F.~Ortega, A.~Hernando, A.~Guti{\'{e}}rrez, Recommender systems survey, Knowledge-Based Systems 46 (2013) 109--132.
\newblock \href {https://doi.org/10.1016/j.knosys.2013.03.012} {\path{doi:10.1016/j.knosys.2013.03.012}}.

\bibitem{Herlocker2002}
J.~Herlocker, J.~A. Konstan, J.~Riedl, An empirical analysis of design choices in neighborhood-based collaborative filtering algorithms, Information Retrieval 5 (2002) 287--310.
\newblock \href {https://doi.org/10.1023/A:1020443909834} {\path{doi:10.1023/A:1020443909834}}.

\bibitem{Khusro2016}
S.~Khsuro, Z.~Ali, I.~Ullah, Recommender systems: Issues, challenges, and research opportunities, in: Proceedings of the 7th International Conference on Information Science and Applications, ICISA 2016, Springer Science+Business Media, Berlin/Heidelberg, Germany, 2016, pp. 1179--1189.
\newblock \href {https://doi.org/10.1007/978-981-10-0557-2\_112} {\path{doi:10.1007/978-981-10-0557-2\_112}}.

\bibitem{Koren2009a}
Y.~Koren, R.~Bell, C.~Volinsky, Matrix factorization techniques for recommender systems, Computer 42~(8) (2009) 30--37.
\newblock \href {https://doi.org/10.1109/MC.2009.263} {\path{doi:10.1109/MC.2009.263}}.

\bibitem{Mikolov2013b}
T.~Mikolov, I.~Sutskever, K.~Chen, G.~Conrado, J.~Dan, Distributed representations of words and phrases and their compositionality, in: Proceedings of the 26th International Conference on Neural Information Processing Systems, NIPS 2013, Curran Associates Inc., Red Hook, NY, USA, 2013, pp. 3111--3119.
\newblock \href {https://doi.org/10.5555/2999792.2999959} {\path{doi:10.5555/2999792.2999959}}.

\bibitem{Grbovic2015}
M.~Grbovic, V.~Radosavljevic, N.~Djuric, N.~Bhamidipati, J.~Savla, V.~Bhagwan, D.~Sharp, E-commerce in your inbox: Product recommendations at scale, in: Proceedings of the 21th ACM SIGKDD International Conference on Knowledge Discovery and Data Mining, KDD '15, Association for Computing Machinery, New York, NY, USA, 2015, pp. 1809--1818.
\newblock \href {https://doi.org/10.1145/2783258.2788627} {\path{doi:10.1145/2783258.2788627}}.

\bibitem{Barkan2016}
O.~Barkan, N.~Koenigstein, {Item2Vec}: Neural item embedding for collaborative filtering, in: IEEE 26th International Workshop on Machine Learning for Signal Processing, MLSP 2016, IEEE, Piscataway, NJ, USA, 2016, pp. 1--6.
\newblock \href {https://doi.org/10.1109/MLSP.2016.7738886} {\path{doi:10.1109/MLSP.2016.7738886}}.

\bibitem{Kang2018}
W.-C. Kang, J.~McAuley, Self-attentive sequential recommendation, in: Proceedings of the 2018 IEEE International Conference on Data Mining, ICDM 2018, 2018, pp. 197--206.
\newblock \href {https://doi.org/10.1109/ICDM.2018.00035} {\path{doi:10.1109/ICDM.2018.00035}}.

\bibitem{Sun2019}
F.~Sun, J.~Liu, J.~Wu, C.~Pei, X.~Lin, W.~Ou, P.~Jiang, {BERT4Rec}: Sequential recommendation with bidirectional encoder representations from transformer, in: Proceedings of the 28th ACM International Conference on Information and Knowledge Management, CIKM '19, Association for Computing Machinery, New York, NY, USA, 2019, p. 1441–1450.
\newblock \href {https://doi.org/10.1145/3357384.3357895} {\path{doi:10.1145/3357384.3357895}}.

\bibitem{Xie2022}
X.~Xie, F.~Sun, Z.~Liu, S.~Wu, J.~Gao, J.~Zhang, B.~Ding, B.~Cui, Contrastive learning for sequential recommendation, in: Proceedings of the IEEE 38th International Conference on Data Engineering, ICDE 2022, IEEE Computer Society, Washington, D.C., USA, 2022, pp. 1--11.
\newblock \href {https://doi.org/10.1109/ICDE53745.2022.00099} {\path{doi:10.1109/ICDE53745.2022.00099}}.

\bibitem{Fu2017}
P.~FU, J.~hua LV, S.~long MA, B.~jie LI, {Attr2vec:} a neural network based item embedding method, in: Proceedings of the 2nd International Conference on Computer, Mechatronics and Electronic Engineering, CMEE 2017, DEStech Publications, Lancaster, PA, USA, 2017, pp. 300--307.
\newblock \href {https://doi.org/10.12783/dtcse/cmee2017/19993} {\path{doi:10.12783/dtcse/cmee2017/19993}}.

\bibitem{Hasanzadeh2022}
S.~Hasanzadeh, S.~M. Fakhrahmad, M.~Taheri, Review-based recommender systems: A proposed rating prediction scheme using word embedding representation of reviews, The Computer Journal 65~(2) (2022) 345--354.
\newblock \href {https://doi.org/10.1093/comjnl/bxaa044} {\path{doi:10.1093/comjnl/bxaa044}}.

\bibitem{Zarzour2019}
H.~Zarzour, Z.~A. Al-Sharif, Y.~Jararweh, {RecDNNing}: a recommender system using deep neural network with user and item embeddings, in: Proceedings of the 10th International Conference on Information and Communication Systems, ICICS 2019, IEEE, New York, NY, USA, 2019, pp. 99--103.
\newblock \href {https://doi.org/10.1109/IACS.2019.8809156} {\path{doi:10.1109/IACS.2019.8809156}}.

\bibitem{Zhang2016a}
F.~Zhang, N.~J. Yuan, D.~Lian, X.~Xie, W.-Y. Ma, Collaborative knowledge base embedding for recommender systems, in: Proceedings of the 22nd ACM SIGKDD International Conference on Knowledge Discovery and Data Mining, KDD '16, Association for Computing Machinery, New York, NY, USA, 2016, pp. 353--362.
\newblock \href {https://doi.org/10.1145/2939672.2939673} {\path{doi:10.1145/2939672.2939673}}.

\bibitem{Camacho2018}
J.~Camacho-Collados, M.~T. Pilehvar, From word to sense embeddings: A survey on vector representations of meaning, Journal of Artificial Intelligence Research 63~(1) (2018) 743--788.
\newblock \href {https://doi.org/10.1613/jair.1.11259} {\path{doi:10.1613/jair.1.11259}}.

\bibitem{Vasile2016}
F.~Vasile, E.~Smirnova, A.~Conneau, Meta-prod2vec: Product embeddings using side-information for recommendation, in: Proceedings of the 10th ACM Conference on Recommender Systems, RecSys '16, Association for Computing Machinery, New York, NY, USA, 2016, pp. 225--232.
\newblock \href {https://doi.org/10.1145/2959100.2959160} {\path{doi:10.1145/2959100.2959160}}.

\bibitem{Greenstein2017}
A.~Greenstein-Messica, L.~Rokach, M.~Friedman, Session-based recommendations using item embedding, in: Proceedings of the 22nd International Conference on Intelligent User Interfaces, IUI '17, Association for Computing Machinery, New York, NY, USA, 2017, pp. 629--633.
\newblock \href {https://doi.org/10.1145/3025171.3025197} {\path{doi:10.1145/3025171.3025197}}.

\bibitem{Tang2018}
J.~Tang, K.~Wang, Personalized top-n sequential recommendation via convolutional sequence embedding, in: Proceedings of the 11th ACM International Conference on Web Search and Data Mining, WSDM '18, Association for Computing Machinery, New York, NY, USA, 2018, pp. 565--573.
\newblock \href {https://doi.org/10.1145/2939672.2939673} {\path{doi:10.1145/2939672.2939673}}.

\bibitem{Hidasi2016}
B.~Hidasi, A.~Karatzoglou, L.~Baltrunas, D.~Tikk, Session-based recommendations with recurrent neural networks, in: Proceedings of the International Conference on Learning Representations, ICLR 2016, OpenReview, Amherst, MA, USA, 2016, pp. 1--10.

\bibitem{Tan2016}
Y.~K. Tan, X.~Xu, Y.~Liu, Improved recurrent neural networks for session-based recommendations, in: Proceedings of the 1st Workshop on Deep Learning for Recommender Systems, DLRS 2016, Association for Computing Machinery, New York, NY, USA, 2016, pp. 17–--22.
\newblock \href {https://doi.org/10.1145/2988450.2988452} {\path{doi:10.1145/2988450.2988452}}.

\bibitem{Sidana2021}
S.~Sidana, M.~Trofimov, O.~Horodnytskyi, C.~Laclau, Y.~Maximov, M.-R. Amini, User preference and embedding learning with implicit feedback for recommender systems, Data Mining and Knowledge Discovery 35 (2021) 568--592.
\newblock \href {https://doi.org/10.1007/s10618-020-00730-8} {\path{doi:10.1007/s10618-020-00730-8}}.

\bibitem{Grbovic2018}
M.~Grbovic, H.~Cheng, Real-time personalization using embeddings for search ranking at {Airbnb}, in: Proceedings of the 24th ACM SIGKDD International Conference on Knowledge Discovery and Data Mining, KDD '18, Association for Computing Machinery, New York, NY, USA, 2018, pp. 311--320.
\newblock \href {https://doi.org/10.1145/3219819.3219885} {\path{doi:10.1145/3219819.3219885}}.

\bibitem{Valcarce2019}
D.~Valcarce, A.~Landin, J.~Parapar, Álvaro Barreiro, Collaborative filtering embeddings for memory-based recommender systems, Engineering Applications of Artificial Intelligence 85 (2019) 347--356.
\newblock \href {https://doi.org/10.1016/j.engappai.2019.06.020} {\path{doi:10.1016/j.engappai.2019.06.020}}.

\bibitem{Collins2019}
A.~Collins, J.~Beel, Document embeddings vs. keyphrases vs. terms for recommender systems: A large-scale online evaluation, in: Proceedings of the 2019 ACM/IEEE Joint Conference on Digital Libraries, JCDL 2019, IEEE, New York, NY, USA, 2019, pp. 130--133.
\newblock \href {https://doi.org/10.1109/JCDL.2019.00027} {\path{doi:10.1109/JCDL.2019.00027}}.

\bibitem{Deng2022}
Y.~Deng, Recommender systems based on graph embedding techniques: A review, IEEE Access 10 (2022) 5158--51633.
\newblock \href {https://doi.org/10.1109/ACCESS.2022.3174197} {\path{doi:10.1109/ACCESS.2022.3174197}}.

\bibitem{Liu2019}
P.~Liu, L.~Zhang, J.~A. Gulla, Real-time social recommendation based on graph embedding and temporal context, International Journal of Human-Computer Studies 121 (2019) 58--72.
\newblock \href {https://doi.org/10.1016/j.ijhcs.2018.02.008} {\path{doi:10.1016/j.ijhcs.2018.02.008}}.

\bibitem{Wei2023}
Y.~Wei, X.~Liu, Y.~Ma, X.~Wang, L.~Nie, T.-S. Chua, Strategy-aware bundle recommender system, in: Proceedings of the 46th International ACM SIGIR Conference on Research and Development in Information Retrieval, SIGIR '23, Association for Computing Machinery, New York, NY, USA, 2023, pp. 1198--1207.
\newblock \href {https://doi.org/10.1145/3539618.3591771} {\path{doi:10.1145/3539618.3591771}}.

\bibitem{Liu2022}
J.~Liu, C.~Shi, C.~Yang, Z.~Lu, P.~S. Yu, A survey on heterogeneous information network based recommender systems: Concepts, methods, applications and resources, AI Open 3 (2022) 40--57.
\newblock \href {https://doi.org/10.1016/j.aiopen.2022.03.00} {\path{doi:10.1016/j.aiopen.2022.03.00}}.

\bibitem{Yu2013}
X.~Yu, X.~Ren, Y.~Sun, B.~Sturt, U.~Khandelwal, Q.~Gu, B.~Norick, J.~Han, Recommendation in heterogeneous information networks with implicit user feedback, in: Proceedings of the 7th ACM Conference on Recommender Systems, RecSys '13, Association for Computing Machinery, New York, NY, USA, 2013, p. 347–350.
\newblock \href {https://doi.org/10.1145/2507157.2507230} {\path{doi:10.1145/2507157.2507230}}.

\bibitem{Pham2023}
P.~Pham, L.~T.~T. Nguyen, N.-T. Nguyen, W.~Pedrycz, U.~Yun, J.~C.-W. Lin, B.~Vo, An approach to semantic-aware heterogeneous network embedding for recommender systems, IEEE Transactions on Cybernetics 53~(9) (2023) 6027--6040.
\newblock \href {https://doi.org/10.1109/TCYB.2022.3233819} {\path{doi:10.1109/TCYB.2022.3233819}}.

\bibitem{Forouzandeh2024}
S.~Forouzandeh, K.~Berahmand, M.~Rostami, A.~Aminzadeh, M.~Oussalah, {UIFRS-HAN}: User interests-aware food recommender system based on the heterogeneous attention network, Engineering Applications of Artificial Intelligence 135~(108766) (2024) 1–16.
\newblock \href {https://doi.org/10.1016/j.engappai.2024.108766} {\path{doi:10.1016/j.engappai.2024.108766}}.

\bibitem{Gao2023}
C.~Gao, Y.~Zheng, N.~Li, Y.~Li, Y.~Qin, J.~Piao, Y.~Quan, J.~Chang, D.~Jin, X.~He, Y.~Li, A survey of graph neural networks for recommender systems: Challenges, methods, and directions, ACM Transactions on Recommender Systems 1~(1) (2023) 3:1–3:51.
\newblock \href {https://doi.org/10.1145/3568022} {\path{doi:10.1145/3568022}}.

\bibitem{He2020}
X.~He, K.~Deng, X.~Wang, Y.~Li, Y.~Zhang, M.~Wang, {LightGCN}: Simplifying and powering graph convolution network for recommendation, in: Proceedings of the 43rd International ACM SIGIR Conference on Research and Development in Information Retrieval, SIGIR '20, Association for Computing Machinery, New York, NY, USA, 2020, pp. 639--–648.
\newblock \href {https://doi.org/10.1145/3397271.3401063} {\path{doi:10.1145/3397271.3401063}}.

\bibitem{Bendada2023}
W.~Bendada, G.~Salha-Galvan, R.~Hennequin, T.~Bouabça, T.~Cazenave, On the consistency of average embeddings for item recommendation, in: Proceedings of the 17th ACM Conference on Recommender Systems, RecSys '23, Association for Computing Machinery, New York, NY, USA, 2023, pp. 833--839.
\newblock \href {https://doi.org/10.1145/3604915.3608837} {\path{doi:10.1145/3604915.3608837}}.

\bibitem{Dacrema2019}
M.~F. Dacrema, P.~Cremonesi, D.~Jannach, Are we really making much progress? a worrying analysis of recent neural recommendation approaches, in: Proceedings of the 13th ACM Conference on Recommender Systems, RecSys '19, Association for Computing Machinery, New York, NY, USA, 2019, pp. 101--109.
\newblock \href {https://doi.org/10.1145/3298689.3347058} {\path{doi:10.1145/3298689.3347058}}.

\bibitem{Gladkova2016}
A.~Gladkova, A.~Drozd, Intrinsic evaluations of word embeddings: What can we do better?, in: Proceedings of the 1st Workshop on Evaluating Vector Space Representations for NLP, Association for Computational Linguistics, Stroudsburg, PA, USA, 2016, pp. 36--42.
\newblock \href {https://doi.org/10.18653/v1/W16-2507} {\path{doi:10.18653/v1/W16-2507}}.

\bibitem{Schnabel2015}
T.~Schnabel, I.~Labutov, D.~Mimno, T.~Joachims, Evaluation methods for unsupervised word embeddings, in: Proceedings of the 2015 Conference on Empirical Methods in Natural Language Processing, EMNLP 2015, Association for Computational Linguistics, Stroudsburg, PA, USA, 2000, pp. 298--307.
\newblock \href {https://doi.org/10.18653/v1/D15-1036} {\path{doi:10.18653/v1/D15-1036}}.

\bibitem{Chang2023}
C.~Chang, J.~Zhou, Y.~Weng, X.~Zeng, Z.~Wu, C.-D. Wang, Y.~Tang, {KGTN}: Knowledge graph transformer network for explainable multi-category item recommendation, Knowledge-Based Systems 278 (2023) 110854.
\newblock \href {https://doi.org/10.1016/j.knosys.2023.110854} {\path{doi:10.1016/j.knosys.2023.110854}}.

\bibitem{Siswanto2018}
A.~V. Siswanto, L.~Tjong, Y.~Saputra, Simple vector representations of e-commerce products, in: 2018 International Conference on Asian Language Processing, IALP 2018, IEEE, New York, NY, USA, 2018, pp. 368--372.
\newblock \href {https://doi.org/10.1109/IALP.2018.8629245} {\path{doi:10.1109/IALP.2018.8629245}}.

\bibitem{Hu2008}
Y.~Hu, Y.~Koren, C.~Volinsky, Collaborative filtering for implicit feedback datasets, in: Proceedings of the 8th IEEE International Conference on Data Mining, ICDM '08, IEEE Computer Society, Washington, D.C., USA, 2008, pp. 263--272.
\newblock \href {https://doi.org/10.1109/ICDM.2008.22} {\path{doi:10.1109/ICDM.2008.22}}.

\bibitem{Caselles2018}
H.~Caselles-Dupres, F.~Lesaint, J.~Royo-Letelier, Word2vec applied to recommendation: hyperparameters matter, in: Proceedings of the 12th ACM Conference on Recommender Systems, RecSys '18, Association for Computing Machinery, New York, NY, USA, 2018, pp. 352--356.
\newblock \href {https://doi.org/10.1145/3240323.3240377} {\path{doi:10.1145/3240323.3240377}}.

\bibitem{Deshpande2004}
M.~Deshpande, G.~Karypis, Item-based top-n recommendation algorithms, ACM Transactions on Information Systems 22~(1) (2004) 143--177.
\newblock \href {https://doi.org/10.1145/963770.963776} {\path{doi:10.1145/963770.963776}}.

\bibitem{Strassen1969}
V.~Strassen, Gaussian elimination is not optimal, Numerische Mathematik 13 (1969) 354--–356.
\newblock \href {https://doi.org/10.1007/BF02165411} {\path{doi:10.1007/BF02165411}}.

\bibitem{Rendle2009}
S.~Rendle, C.~Freudenthaler, Z.~Gantner, L.~Schmidt-Thieme, {BPR}: Bayesian personalized ranking from implicit feedback, in: Proceedings of the 25th Conference on Uncertainty in Artificial Intelligence, UAI '09, AUAI Press, Arlington, VA, USA, 2009, pp. 452--461.
\newblock \href {https://doi.org/10.5555/1795114.1795167} {\path{doi:10.5555/1795114.1795167}}.

\bibitem{Shenbin2019}
I.~Shenbin, A.~Alekseev, E.~Tutubalina, V.~Malykh, S.~I. Nikolenko, {RecVAE}: A new variational autoencoder for top-n recommendations with implicit feedback, in: Proceedings of the 13th International Conference on Web Search and Data Mining, WSDM '20, Association for Computing Machinery, New York, NY, USA, 2020, p. 528–536.
\newblock \href {https://doi.org/10.1145/3336191.3371831} {\path{doi:10.1145/3336191.3371831}}.

\bibitem{Sarwar2001}
B.~M. Sarwar, G.~Karypis, J.~A. Konstan, J.~T. Riedl, Item-based collaborative filtering recommendation algorithms, in: Proceedings of the 10th International Conference on World Wide Web, WWW '01, Association for Computing Machinery, New York, NY, USA, 2001, pp. 285--295.
\newblock \href {https://doi.org/10.1145/371920.372071} {\path{doi:10.1145/371920.372071}}.

\bibitem{Sarwar2002}
B.~M. Sarwar, G.~Karypis, J.~Konstan, J.~Riedl, Recommender systems for large-scale e-commerce: Scalable neighborhood formation using clustering, in: Proceedings of The 5th International Conference on Computer and Information Technology, ICCIT 2002, Dhaka, Bangladesh, 2002, pp. 1--6.

\bibitem{Demsar2006}
J.~Dem{\v s}ar, Statistical comparisons of classifiers over multiple data sets, The Journal of Machine Learning Research 7 (2006) 1--30.
\newblock \href {https://doi.org/10.5555/1248547.1248548} {\path{doi:10.5555/1248547.1248548}}.

\bibitem{Pires2024}
P.~R. Pires, B.~B. Rizzi, T.~A. Almeida, Why ignore content? {A} guideline for intrinsic evaluation of item embeddings for collaborative filtering, in: Proceedings of the 30th Brazilian Symposium on Multimedia and the Web, WebMedia '30, Sociedade Brasileira de Computação, Juiz de Fora, Brazil, 2024, pp. 345--354.
\newblock \href {https://doi.org/10.5753/webmedia.2024.243199} {\path{doi:10.5753/webmedia.2024.243199}}.

\bibitem{Koren2009b}
Y.~Koren, Collaborative filtering with temporal dynamics, in: Proceedings of the 15th ACM SIGKDD international conference on Knowledge discovery and data mining, KDD '09, 2009, pp. 447--456.

\bibitem{Pires2021}
P.~R. Pires, A.~C. Pascon, T.~A. Almeida, Time-dependent item embeddings for collaborative filtering, in: Proceedings of the 10th Brazilian Conference on Intelligent Systems, BRACIS '21, Springer Nature, Virtual Event, Brazil, 2021, pp. 309--324.
\newblock \href {https://doi.org/10.1007/978-3-030-91699-2\_22} {\path{doi:10.1007/978-3-030-91699-2\_22}}.

\bibitem{Brody2024}
S.~Brody, S.~Lagziel, Simrec: Mitigating the cold-start problem in sequential recommendation by integrating item similarity (2024).
\newblock \href {http://arxiv.org/abs/2410.22136} {\path{arXiv:2410.22136}}.

\bibitem{Rehurek2010}
R.~{\v R}eh{\r u}{\v r}ek, P.~Sojka, Software framework for topic modelling with large corpora, in: Proceedings of the LREC 2010 Workshop on New Challenges for NLP Frameworks, LREC 2010, European Language Resources Association (ELRA), Paris, France, 2010, pp. 45--50.
\newblock \href {https://doi.org/10.13140/2.1.2393.1847} {\path{doi:10.13140/2.1.2393.1847}}.

\end{thebibliography}





\end{document}